\documentclass[%
 aip,
 amsmath,amssymb,
 reprint,%
]{revtex4-1}

\usepackage{graphicx}
\usepackage{dcolumn}
\usepackage{bm}

\usepackage[utf8]{inputenc}
\usepackage[T1]{fontenc}
\usepackage{etoolbox}
\usepackage[colorlinks=true,allcolors=blue]{hyperref}
\usepackage{epstopdf, epsfig}
\usepackage{newtxtext,newtxmath} 
\usepackage[]{subcaption} 
\usepackage{placeins}

\makeatletter
\def\@email#1#2{%
 \endgroup
 \patchcmd{\titleblock@produce}
  {\frontmatter@RRAPformat}
  {\frontmatter@RRAPformat{\produce@RRAP{*#1\href{mailto:#2}{#2}}}\frontmatter@RRAPformat}
  {}{}
}%
\makeatother
\begin{document}
\preprint{AIP/123-QED}

\title[Phase-Averaged Dynamics  of a Periodically Surging Wind Turbine]{Phase-Averaged Dynamics  of a Periodically Surging Wind Turbine}
\author{Nathaniel J. Wei}
 \affiliation{Graduate Aerospace Laboratories, California Institute of Technology}
\author{John O. Dabiri}%
 \email{jodabiri@caltech.edu}
 \affiliation{Graduate Aerospace Laboratories, California Institute of Technology}
\affiliation{Department of Mechanical and Civil Engineering, California Institute of Technology}%

\date{\today}

\begin{abstract}
The unsteady power generation of a wind turbine translating in the streamwise direction is relevant to floating offshore wind turbines, kite-mounted airborne wind turbines, and other non-traditional wind-energy systems. To study this problem experimentally, measurements of torque, rotor speed, and power were acquired for a horizontal-axis wind turbine actuated in periodic surge motions in a fan-array wind tunnel at the Caltech Center for Autonomous Systems and Technologies (CAST). Experiments were conducted at a diameter-based Reynolds number of $Re_D=6.1\times10^5$ and at tip-speed ratios between 5.2 and 8.8. Sinusoidal and trapezoidal surge-velocity waveforms with maximum surge velocities up to 23\% of the free-stream velocity were tested. A model in the form of a linear ordinary differential equation (first-order in time) was derived to capture the time-resolved dynamics of the surging turbine. Its coefficients were obtained using torque measurements from a stationary turbine, without the need for unsteady calibrations. Its predictions compared favorably with the measured amplitude- and phase-response data. Furthermore, increases in the period-averaged power of up to 6.4\% above the steady reference case were observed in the experiments at high tip-speed ratios and surge velocities, potentially due to unsteady or nonlinear aerodynamic effects. Conversely, decreases in mean power with increased surge velocity at low tip-speed ratios were likely a result of the onset of stall on the turbine blades. These results inform the development of strategies to optimize and control the unsteady power generation of periodically surging wind turbines, and motivate further investigations into the unsteady aerodynamics of wind-energy systems.
\end{abstract}

\maketitle

\section{Introduction}
\label{sec:intro}

New innovations in wind energy technology motivate the study of wind-turbine performance in previously unexplored operational regimes. In particular, while traditional land-based wind turbines are fixed in place, wind-energy systems such as floating offshore wind turbines (FOWTs) and airborne wind turbines undergo streamwise oscillations that may potentially complicate the aerodynamics of these systems. The periodic motions of the turbine rotor in these situations introduce additional dynamics that can affect the power generation of the wind turbines and the fatigue loading on their blades, thereby impacting their contribution to global energy demands. Therefore, this study investigates the dynamics of a periodically surging wind turbine through analytical modeling and laboratory-scale experiments.

\subsection{Current Progress in Surging-Turbine Aerodynamics}

Previous studies have generally considered surge motions typical of wave-driven FOWTs,\cite{gueydon_discussion_2020} which are of increasing relevance as the emerging offshore-wind sector continues to expand. The majority of attention regarding turbine aerodynamics has been focused on time-averaged quantities. Using a model FOWT in a wind tunnel and wave tank, \citeauthor{farrugia_investigating_2014}\cite{farrugia_investigating_2014} found that the time-averaged coefficient of power, $\overline{C_p}$, increased above the steady case by 1\% when oscillations in the turbine were present. A similar increase in $\overline{C_p}$ by 1\% was observed in wind-tunnel experiments by \citeauthor{khosravi_experimental_2015}\cite{khosravi_experimental_2015} and free-vortex wake simulations by \citeauthor{shen_study_2018}\cite{shen_study_2018}. \citeauthor{farrugia_study_2016}\cite{farrugia_study_2016} showed using free-vortex wake simulations that $\overline{C_p}$ increased with surge frequency at tip-speed ratios above the rated value by up to 13.7\%, but decreased with surge frequency at tip-speed ratios below the rated value. Independent simulations by \citeauthor{wen_influences_2017}\cite{wen_influences_2017} yielded similar results. \citeauthor{johlas_floating_2021}\cite{johlas_floating_2021} suggested that the increases in average power with surge velocity can be described by a simple quasi-steady model, where the term ``quasi-steady'' refers to effects for which successive instances in time can be considered as being in independent states of local equilibrium. Since the model is derived from the cubic dependence of power on the incident inflow velocity at the rotor, the relative power gains over the steady value of $C_p$ from surge motions in the upwind direction outweigh the relative losses from surge motions in the downwind direction. Their model agrees well with the time-averaged power results from their numerical simulations of a surging turbine, but the validity of the model has not yet been evaluated over a wide range of surge velocities and operating conditions. A fully characterized, quantitatively accurate explanation for the observed increases in $\overline{C_p}$ thus remains elusive.

The time-resolved dynamics of a turbine in surge have been explored as well. These unsteady dynamics determine the unsteady loads on the turbine and its support structures, and therefore inform the design of FOWT control systems. The presence of fluctuations in turbine thrust, torque, and power at the same frequency as the imposed surge motion is well-documented in the literature.\cite{farrugia_investigating_2014,farrugia_study_2016,shen_study_2018,tran_cfd_2016,wen_influences_2017,johlas_floating_2021} These fluctuations increase in amplitude as the surge frequency is increased.\cite{farrugia_study_2016} \citeauthor{mancini_characterization_2020}\cite{mancini_characterization_2020}, however, showed in wind-tunnel experiments with a surging turbine that the relationship between torque amplitude and surge frequency increases above the prediction of their linear quasi-steady model at high frequencies. They attributed this to mechanical resonance, and not to a breakdown of their linearization of power as a function of the surge velocity or the influence of unsteady aerodynamics. By contrast, torque amplitudes measured in wind-tunnel experiments by \citeauthor{sant_measurements_2015}\cite{sant_measurements_2015} were much lower than those computed by quasi-steady and dynamic-inflow codes.

The phase response of the turbine similarly lacks a single consistent characterization in the literature. The model developed by \citeauthor{johlas_floating_2021}\cite{johlas_floating_2021} predicts that the instantaneous power from the turbine will be in phase with the surge velocity. The linear quasi-steady model of \citeauthor{mancini_characterization_2020}\cite{mancini_characterization_2020} supports the same prediction. Some computational\cite{farrugia_study_2016,tran_cfd_2016,wen_influences_2017,wen_power_2018} and experimental\cite{mancini_characterization_2020} results, however, have shown phase differences in excess of $90^\circ$, while others displayed close to zero phase offset.\cite{micallef_loading_2015,shen_study_2018,wen_power_2018-1} The discrepancies in the literature regarding the amplitude and phase of the torque and power output of surging turbines motivates the current study.

The lack of consensus with respect to mean quantities and their amplitude and phase stems in large part from unanswered questions regarding the relative importance of quasi-steady and unsteady effects. The models of \citeauthor{johlas_floating_2021}\cite{johlas_floating_2021} and \citeauthor{mancini_characterization_2020}\cite{mancini_characterization_2020} can be classified as purely quasi-steady models that neglect unsteady effects. Other models have incorporated unsteady effects directly. For example, \citeauthor{de_vaal_effect_2014}\cite{de_vaal_effect_2014} compared the results of different dynamic inflow models, which include corrections for time-varying inflow velocity and acceleration, that were paired with blade-element momentum (BEM) simulations of a surging turbine. They concluded that these engineering models were capable of capturing global forces on FOWTs in typical offshore conditions. In a different approach, \citeauthor{fontanella_control-oriented_2020}\cite{fontanella_control-oriented_2020} derived a state-space model that maps linearized turbine aerodynamics and wave dynamics to the time derivatives of the kinematic parameters of the turbine. The model was shown to perform well both in simulations and as the basis for control systems.\cite{fontanella_model-based_2021} In addition to these models, others have suggested various unsteady flow phenomena that could influence the turbine dynamics. For instance, several of the aforementioned studies have considered the effects of airfoil stall, particularly at the blade root, on time-averaged and fluctuating quantities.\cite{farrugia_study_2016,tran_cfd_2016,wen_influences_2017,wen_power_2018} In addition to blade stall, \citeauthor{sebastian_characterization_2013}\cite{sebastian_characterization_2013} postulated the formation of unsteady recirculation regions in or downstream of the rotor plane during turbine surge, as a result of slip-stream violations. Furthermore, \citeauthor{wen_power_2018}\cite{wen_power_2018} attributed the phase differences observed in their simulations to added-mass effects, blade-wake interactions, and unsteady aerodynamics. These unsteady flow phenomena may affect the structure, dynamics, and recovery of the wake of a surging turbine.\cite{rockel_wake_2016,tran_cfd_2016,bayati_wind_2017,lee_effects_2019,kopperstad_aerodynamic_2020,rezaeiha_wake_2021} However, it still remains to be seen which (if any) of these unsteady effects must be accounted for in a model to capture the torque and power production of real surging turbines, or whether existing quasi-steady models are sufficient for this purpose.

Lastly, since nearly all existing work on surging-turbine aerodynamics has been conducted in the context of surge oscillations typical of FOWTs under normal operating conditions, the dynamics of wind turbines surging through larger amplitudes or higher frequencies remain relatively unexplored. Larger surge oscillations would be relevant not only to FOWTs in more extreme conditions, but also to airborne wind turbines mounted to aircraft or crosswind kites.\cite{cherubini_airborne_2015} Crosswind kites generally fly through large periodic orbits with length scales much greater than the size of the aircraft itself.\cite{jonkman_google_2021} Turbines mounted to these kites would therefore undergo surge motions at amplitudes far larger than those experienced by FOWTs. In addition, \citeauthor{dabiri_theoretical_2020}\cite{dabiri_theoretical_2020} recently suggested that streamwise unsteadiness could be leveraged to increase the efficiency of wind-energy systems above the theoretical steady limit. Since increases in time-averaged power have already been observed at the relatively low levels of unsteady motion typical of FOWTs, an investigation of higher surge amplitudes and frequencies could provide insights toward the practical realization of these theoretical efficiency gains.

\subsection{Research Objectives}

This study aims to address several open questions regarding the time-resolved dynamics of a wind turbine in surge. The amplitude and phase of torque and power relative to the surge motions are investigated in wind-tunnel experiments. Trends in the data are parameterized by a model that accounts for quasi-steady aerodynamic torques and unsteady generator torques. The model is first-order in time and linear in the turbine surge velocity and rotor speed; thus, it shall henceforth be referred to as a first-order linear model. An important feature of the model is that its coefficients can be computed from measurements obtained under steady conditions; no data from actual surge tests are required to obtain time-resolved torque and power predictions. The experiments span higher levels of unsteadiness than previous studies in the literature, with scaled amplitudes up to $A^* = A/D = 0.51$ and nondimensional surge velocities up to $u^* = fA/u_1 = 0.23$, where $A$ is the surge amplitude, $f$ is the surge frequency in radians per second, $u_1$ is the free-stream velocity, and $D$ is the turbine diameter. By contrast, the highest values reported in the literature are $A^* = 0.13$ (\citeauthor{tran_cfd_2016}\cite{tran_cfd_2016}) and $u^* = 0.42$ (\citeauthor{wen_power_2018}\cite{wen_power_2018}) in simulations, and $A^* = 0.15$ (\citeauthor{sant_measurements_2015}\cite{sant_measurements_2015}) and $u^* = 0.079$ (\citeauthor{mancini_characterization_2020}\cite{mancini_characterization_2020}) in experiments. The findings in this study may thus be generalized to FOWTs operating under extreme conditions, as well as novel airborne wind-energy systems and other emergent technologies. The combined analytical and experimental results presented in this work provide a foundation upon which questions regarding the influence of unsteadiness and nonlinearity, including the dependence of the mean torque and power on surge kinematics, may be more comprehensively investigated in future work.

The paper is structured as follows. In Section \ref{sec:model}, a first-order linear model is derived that enables a disambiguation between aerodynamic and generator torques. Its amplitude and phase characteristics are also analyzed. In Section \ref{sec:methods}, the experimental apparatus is described, and methods for computing the coefficients of the analytical model from measurements in steady conditions are given. Phase-averaged results from experiments with sinusoidal and trapezoidal surge-velocity waveforms are presented in Section \ref{sec:results}, and the results are compared with model predictions. Finally, a discussion regarding model capabilities and limitations, nonlinear and unsteady effects, and application to full-scale wind turbines is provided in Section \ref{sec:discussion}.

\section{Analytical Model}
\label{sec:model}

In this section, we derive a model for the torque generated by a surging horizontal-axis turbine from an ordinary differential equation that is first-order in time and linear in the turbine surge velocity and rotor speed (or rotation rate). We linearize the aerodynamic torque with respect to the inflow velocity and rotor speed, and combine it with a model for the generator torque to obtain a differential equation for the rotor speed of the turbine. We then derive transfer functions in the frequency domain to characterize the amplitude and phase relative to the surge-velocity waveform of the aerodynamic and generator torque. A notable advantage of this model is that the model coefficients can be extracted directly from torque and rotation-rate measurements of the turbine in steady conditions (i.e.\ without surge motions); these methods will be described in Section \ref{sec:constants} for the turbine used in these experiments.

\subsection{Aerodynamic Torque Model}
\label{sec:aeroTorque}

A first-order linear model for the aerodynamic torque can be derived using a local linearization with respect to the inflow velocity and rotor speed:

\begin{equation}
\tau_{aero} \approx \tau_0 + \frac{\partial \tau}{\partial u}\bigg\rvert_{u=u_1,\omega=\overline{\omega}} (u-u_1) + \frac{\partial \tau}{\partial \omega}\bigg\rvert_{u=u_1,\omega=\overline{\omega}} (\omega-\overline{\omega}),
\label{eqn:tauAero0}
\end{equation}

\noindent where $u = u_1 + U(t)$ is the instantaneous inflow velocity relative to the turbine, $U(t)$ is the turbine surge velocity in a stationary frame of reference, $\omega$ is the rotor speed, and $\tau_0$ is the steady aerodynamic torque, i.e.\ the mean torque measured on a stationary turbine at a wind speed of $u_1$. In this work, bars denote time averages over a single streamwise-motion oscillation period for time-dependent variables in the case of unsteady streamwise motion, while the subscript $0$ denotes the value of a variable in the reference case corresponding to a steady inflow at speed $u_1$. We then define the performance coefficients

\begin{equation}
K_\ell = \frac{\partial \tau}{\partial u}\big\rvert_{u=u_1,\omega=\overline{\omega}}
\label{eqn:KL}
\end{equation}
\noindent and 
\begin{equation}
K_d = -\frac{1}{R}\frac{\partial \tau}{\partial \omega}\big\rvert_{u=u_1,\omega=\overline{\omega}},
\label{eqn:Kd}
\end{equation}
\noindent where $R$ is the radius of the turbine. These coefficients qualitatively correspond to lift and drag terms in a blade-element expression for aerodynamic torque. Values for these constants can be obtained empirically from measurements of the turbine torque taken with a stationary turbine over a range of wind speeds and loading conditions (cf.\ Section \ref{sec:constants}). Simplifying the above expression yields the following model:

\begin{equation}
\tau_{aero} \approx K_\ell U - K_d R (\omega-\overline{\omega}) + \tau_0.
\label{eqn:tauAero}
\end{equation}

The accuracy of this aerodynamic model depends on whether $\tau$ is sufficiently linear in $u$ and $\omega$ in the neighborhood of the steady operating condition ($u=u_1$ and $\omega=\overline{\omega}$). Since the model is inherently quasi-steady, its accuracy will also depend on whether any unsteady effects such as dynamic stall on the turbine blades are present.

\subsection{Generator Torque Model}
\label{sec:genTorque}

The torque applied by the generator ($\tau_{gen}$) in opposition to the aerodynamic torque ($\tau_{aero}$) represents the torque that is converted to usable power at each instant in time. It thus also represents the mechanically measurable torque on the turbine shaft ($\tau_{meas}$). It is important to note that the generator torque is not necessarily equal to the aerodynamic torque in the case of unsteady rotation, with any difference between the two inducing a change in the angular velocity of the rotor.

The equations of motion for a permanent-magnet generator are identical in principle to those for a permanent-magnet motor, which is frequently modeled as a first-order ordinary differential equation in time:\cite{concordia_synchronous_1951}

\begin{equation}
\tau_{gen} = \tau_{meas} = K_2 \frac{d\omega}{dt} + K_1 \omega + K_0,
\label{eqn:tauGen}
\end{equation}

\noindent where $K_2$ is the moment of inertia of the generator about its rotational axis, $K_1$ is the generator constant, and $K_0$ is an empirical zero-speed offset. Since the generator torque is proportional to the current through the generator coils, $K_1$ and $K_0$ scale inversely with the resistive load applied to the generator.\cite{concordia_synchronous_1951} Hence, a higher resistive load applied to the generator corresponds to a lower physical load on the turbine. This formulation assumes that the generator is directly driven by the turbine; a gear-ratio scaling could be incorporated to map the rotor speed to the generator rotation rate for turbines with gearboxes.

It should also be noted that the generator model in its current form does not include any effects of control, such as tip-speed ratio control systems that are typically present in utility-scale wind turbines. For the purposes of this study, the use of a direct-drive generator with fixed resistive loading and no speed control simplify the modeling of the turbine dynamics and subsequent model validation against experimental data. However, the linear form of the model means that linear or linearized tip-speed ratio controllers can readily be incorporated using classical analytical techniques.

\subsection{Governing Equation and its Transfer Functions}
\label{sec:fullModel}

The dynamics of a turbine under the influence of competing aerodynamic and generator torques are given by the swing equation,\cite{stevenson_power_1994}

\begin{equation}
J \frac{d\omega}{dt} = \tau_{aero} - \tau_{gen},
\label{eqn:swingEqn}
\end{equation}

\noindent where $J$ is the moment of inertia of the turbine, its shaft assembly, and the generator about the axis of rotation, and is thus in practice much larger than $K_2$. Deviations of the instantaneous aerodynamic torque away from equilibrium, if not immediately matched by the generator torque, will lead to a change in the rotor speed until the generator torque overcomes inertia and restores the torque balance. This implies that the torque measured by a torque transducer, and consequently the power measured from the generator, will lag behind the aerodynamic torque.

Substituting Equations \ref{eqn:tauAero} and \ref{eqn:tauGen} into the above relation, we arrive at the equation of motion

\begin{equation}
J \frac{d\omega}{dt} = \left(K_\ell U - K_d R (\omega-\overline{\omega}) + \tau_0 \right) - \left(K_2 \frac{d\omega}{dt} + K_1 \omega + K_0\right).
\label{eqn:modelPrelim}
\end{equation}

\noindent In the limit of equilibrium, in which all time-derivatives are zero, the steady aerodynamic torque $\tau_0$ must equal the generator torque, i.e.\ $\tau_0 = K_1 \overline{\omega} + K_0$. We can therefore simplify the model into a more informative form:

\begin{equation}
J \frac{d\omega}{dt} = K_\ell U - K_d R (\omega-\overline{\omega}) - K_2 \frac{d\omega}{dt} - K_1 (\omega-\overline{\omega}).
\label{eqn:model}
\end{equation}

The resulting model is conceptually similar to that of \citeauthor{fontanella_control-oriented_2020}\cite{fontanella_control-oriented_2020}, though in this work the linearization for the aerodynamic torque is obtained differently and the generator torque is an output, rather than an input, to the system. An additional benefit of the model in Equation \ref{eqn:model} is that it requires no data from unsteady surge experiments to make time-resolved predictions, since all of its coefficients can be computed either from measurements in steady flow or from geometric properties of the turbine and generator. The model has the form of a linear time-invariant (LTI) system, which allows transfer functions of the aerodynamic and measured torques to be computed in order to quantify the phase and amplitude behavior of the system. Taking the Laplace transform of Equation \ref{eqn:model} with respect to an arbitrary surge velocity $U$ (input) and the resulting rotor speed $\omega$ (output) yields the transfer function

\begin{equation}
\frac{\omega(s)}{U(s)} = \frac{K_\ell}{(J+K_2)s + K_1 + K_d R}.
\label{eqn:tfRot}
\end{equation}

This transfer function has the form of a first-order low-pass filter with critical frequency $f_c = \frac{K_1 + K_d R}{J + K_2}$. Using this transfer function, we can also derive transfer functions for the aerodynamic and generator torques:

\begin{equation}
\frac{\tau_{aero}(s)}{U(s)} = K_\ell - K_d R \frac{\omega(s)}{U(s)} = K_\ell \frac{(J+K_2)s+K_1}{(J+K_2)s+K_1+K_d R}
\label{eqn:tfAero}
\end{equation}

\noindent and

\begin{equation}
\frac{\tau_{gen}(s)}{U(s)} = (K_2 s + K_1) \frac{\omega(s)}{U(s)} = K_\ell \frac{K_2 s+K_1}{(J+K_2)s+K_1+K_d R},
\label{eqn:tfMeas}
\end{equation}

\noindent which share the same critical frequency $f_c$. The frequency response can be computed from these transfer functions using the imaginary part of the Laplace variable $s$, i.e.\ $\operatorname{Im}(s)=f$. Phase and amplitude predictions from the model can thus be obtained analytically, and the mean torque is given by the steady-state value $\tau_0$. Power can then be computed as $\mathcal{P} = \tau\omega$. The linear form of the model dictates that, for periodic surge motions with zero net displacement, the period-averaged mean torque and power are not functions of the surge motions. According to this model, then, unsteady surge motions will not affect the period-averaged power generation of the turbine. The model thus forgoes the ability to predict time-averaged quantities in favor of an analytical formulation of the time-resolved turbine dynamics. The consequences of this tradeoff will be discussed in Section \ref{sec:discussion_discrepancies}.

The transfer functions suggest that the relevant nondimensional parameters for the surge dynamics are the nondimensional surge velocity, $u^* = fA/u_1$, and the normalized surge frequency, $f^* = f/f_c$. The analysis suggests that the amplitude of the unsteady torque oscillations scales directly with $u^*$, with a frequency dependence dictated by $f^*$. Either the reduced frequency $k = fD/u_1$ or the nondimensional surge amplitude $A^* = A/D$ would complete the nondimensional parameterization by including the length scale of the turbine, but in contrast to suggestions in the literature,\cite{bayati_wind_2017,wen_influences_2017} these parameters do not appear to follow directly from the transfer-function formulation of the model.

\section{Experimental Methods}
\label{sec:methods}

In this section, the experimental apparatus used to study the torque and power production of a wind turbine in periodic surge motions is described. First, the wind tunnel and turbine apparatus are described. Then, the parameter space explored in these experiments is presented, and the experimental procedure is outlined. Finally, methods for empirically determining values for the scaling coefficients of the analytical model derived in the previous section and an overview of sources of uncertainty are provided.

\subsection{Experimental Apparatus}
\label{sec:cast}

\begin{figure*}
	\includegraphics[width=0.7\textwidth]{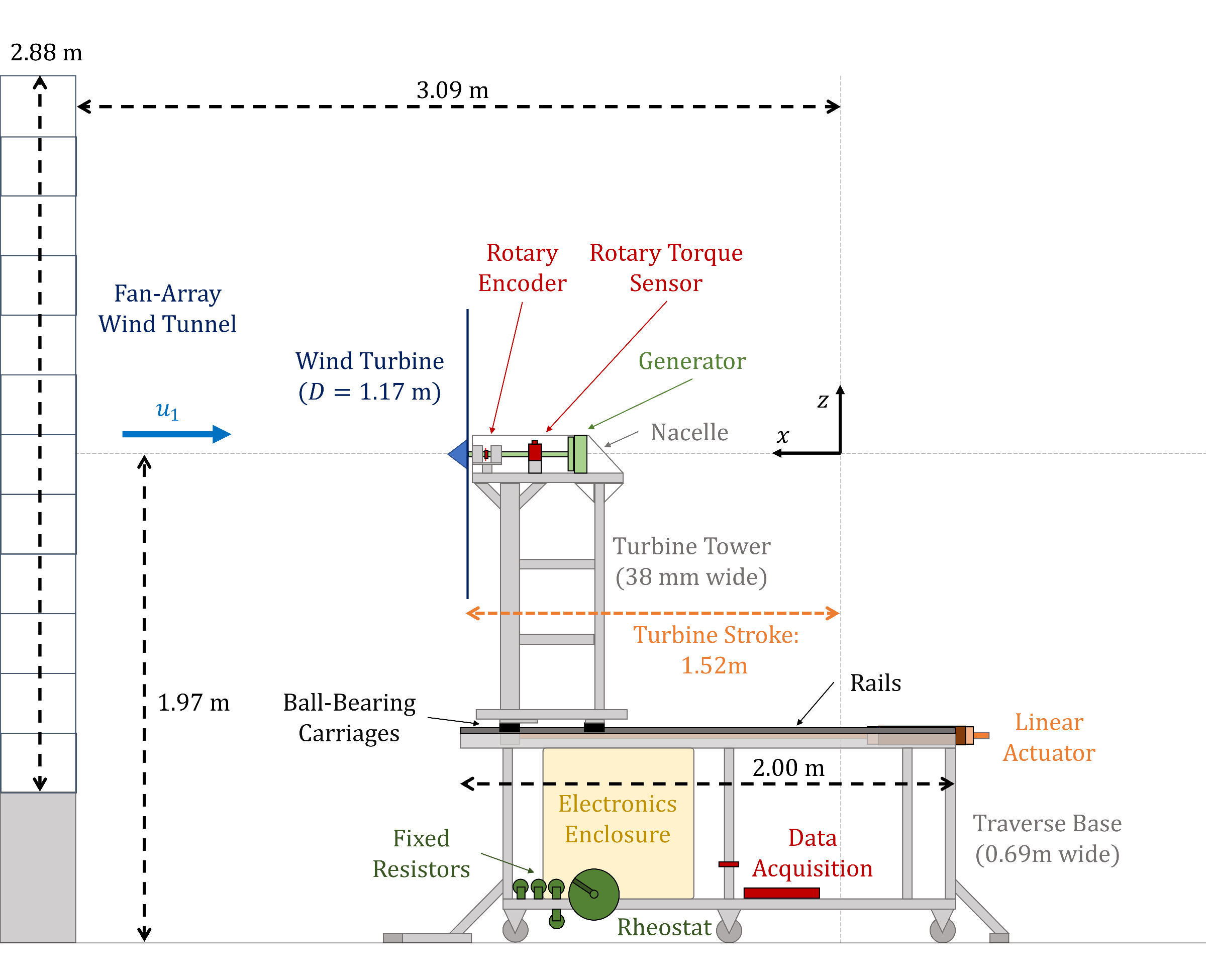}
	\caption{Schematic of the experimental apparatus, including fan-array wind tunnel (left) and surging turbine (center-right). The turbine is illustrated at its maximum upstream position relative to the origin. The estimated blockage of the swept area of the turbine and all support structures, relative to the fan-array surface area, is 14\%.}
	\label{fig:setup}
\end{figure*}

Experiments were conducted in a large open-circuit fan-array wind tunnel at the Caltech Center for Autonomous Systems and Technologies (CAST). The fan array was composed of 2,592 computer fans arranged in two counter-rotating layers within a $2.88 \times 2.88$ m frame, mounted 0.61 m above the floor of the facility (cf.\ Figure \ref{fig:setup}). The open-air test section downstream of the fan array vented directly to the atmosphere, while the other three sides and ceiling of the arena were enclosed with walls or awnings to mitigate the effects of atmospheric disturbances. The experiments in this study were carried out at a free-stream velocity of $u_1 = 8.06 \pm 0.16$ $\rm{ms^{-1}}$, corresponding to a diameter-based Reynolds number of approximately $Re_D = 6.13\times10^5$. The relevance of this study to Reynolds numbers typical of utility-scale wind turbines is discussed in Section \ref{sec:discussion_extensions}. The turbulence intensity in the tunnel, represented by the standard deviation of the velocity fluctuations normalized by the average streamwise velocity, was measured to be $2.20\pm0.17$\%. These measurements were obtained with an ultrasonic anemometer (Campbell Scientific CSAT3B) placed at the hub height and streamwise zero position of the wind turbine. Because the facility was exposed to the atmosphere and experiments were conducted over a range of atmospheric conditions and times of day, temperature and relative-humidity readings were recorded with measurement precisions of $\pm1^\circ$C and $\pm5$\% from a portable weather station (Taylor Precision Products model 1731) so that the air density could be calculated accordingly.

The turbine apparatus was constructed on an aluminum frame (80/20 1515 T-slotted profile) that was bolted to the floor and secured with sandbags. The frame was 2.00 m long, 0.69 m wide, and 0.87 m tall. Two 2-m long rails with two ball-bearing carriages each (NSK NH252000AN2PCZ) were mounted on top of the frame, parallel to the streamwise direction and spaced 0.65 m apart in the cross-stream direction. A traverse was mounted on the ball-bearing carriages, which supported a 0.99-m tall, 0.038-m wide turbine tower. The wind-turbine shaft assembly was placed on top of this tower at a hub height of 1.97 m above the floor. The origin of the surge motions of the turbine was located 3.09 m downstream of the fan array, and the rails afforded a maximum surge stroke of 1.52 m upstream of the origin. A schematic of the apparatus and its position relative to the wind tunnel is given in Figure \ref{fig:setup}.

A three-bladed horizontal-axis wind turbine (Primus Wind Power AIR Silent X) with a rotor diameter of $D = 1.17$ m and hub diameter of 0.127 m was attached to a 25.4-mm diameter steel shaft supported by two axially mounted shaft bearings (Sealmaster NP-16T). The blade chord ranged from 100 mm at the root to 32 mm at the tip. The blades were constructed from a laminated carbon-fiber composite.  A rotary encoder (US Digital EM2-2-4096-I) with a resolution of 4096 counts per revolution was attached between the shaft bearings. A rotary torque transducer (FUTEK TRS300, 20 Nm torque rating) was connected to the turbine shaft on one end and to the turbine generator (Primus Wind Power AIR 30, 48V) by means of a floating 19.0-mm diameter steel shaft on the other. The shaft assembly was surrounded by a 4.76-mm thick acrylic nacelle ($0.610 \times 0.152 \times 0.152$ m) with a slanted rear section intended to reduce bluff-body separation effects in the turbine wake. The estimated blockage of the swept area of the turbine and all support structures, relative to the fan-array surface area, was 14\%.

The load on the turbine was controlled electrically with resistors. The three-phase alternating current produced by the generator was converted to DC by a bridge rectifier (Comchip Technology SC50VB80-G) and passed to a bank of resistors. Different combinations of fixed 10-Ohm resistors (TE Connectivity TE1000B10RJ) in series or parallel and an 8-Ohm rheostat (Ohmite RRS8R0E) were used to achieve a range of loading conditions that spanned the operational envelope of the turbine. An emergency short-circuit switch built from a solid-state relay (Crydom D1D12) and a 12-A fuse (Schurter 4435.0368) were included in the circuit for safety purposes.

The turbine traverse was driven in the streamwise direction by a piston-type magnetic linear actuator (LinMot PS10-70x320U-BL-QJ). Its sliding cylinder was attached to the traverse at rail height and its stator was mounted to the downstream end of the frame. The motions of the traverse were controlled by a servo driver (LinMot E1450-LU-QN-0S) with an external position sensor (LinMot MS01-1/D) mounted along one of the two rails for increased repeatability. Power for the system was provided by a step-up transformer (Maddox MIT-DRY-188). Motion profiles were loaded onto the driver, and a TTL pulse was used to start each successive motion period. In these experiments, the maximum surge velocity was $U = 2.40$ $\rm{ms^{-1}}$ and the maximum surge acceleration was $\frac{dU}{dt} = 23.7$ $\rm{ms^{-2}}$, while the average absolute position error was 0.68 mm and the average absolute velocity error was $8.33\times10^{-3}$ $\rm{ms^{-1}}$.

Data were collected from the rotary encoder and torque transducer using a data acquisition card (National Instruments USB-6221), as well as an amplifier (FUTEK IAA100) for the raw torque voltage signals. Data collection occurred at a sampling rate of 1 kHz. A LabVIEW control program coordinated data collection and triggering for the linear actuator. It was also used to adjust the resistance of the rheostat remotely via a stepper motor (Sparkfun ROB-13656). After each experiment, the program converted the raw voltage signals from the amplifier to dimensional values by interpolating between calibrated points, and it performed a fourth-order central-difference scheme on the angular measurements from the encoder to obtain a rotation-rate signal. Furthermore, the measured torque signals were filtered using a sixth-order Butterworth filter with a cutoff frequency of 100 Hz to reduce the influence of electrical noise.

\subsection{Experimental Parameters}
\label{sec:params}

\begin{figure}
	\centering
	\includegraphics[width=\columnwidth]{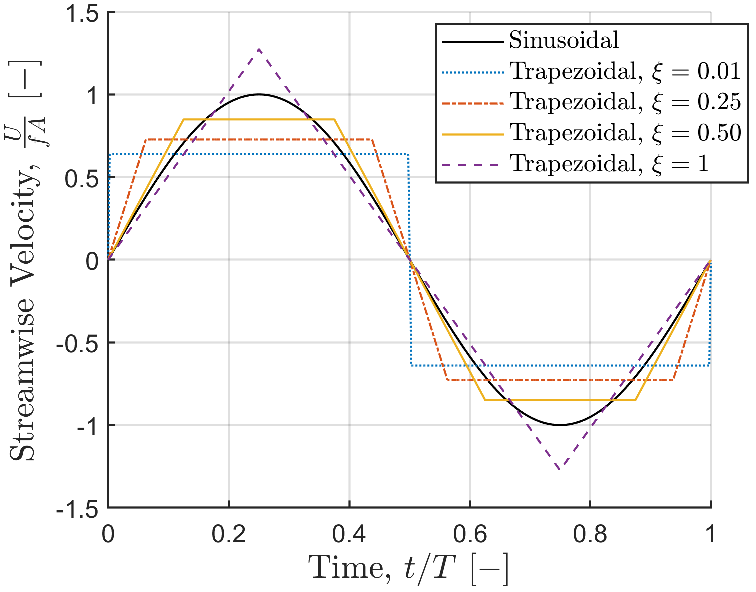}
	\caption{Surge-velocity waveforms used in these experiments. Sinusoidal profiles and four types of trapezoidal profiles (parameterized by $\xi$) were tested.}
	\label{fig:waveforms}
\end{figure}

The apparatus described in the previous section was used to investigate the unsteady torque and power production of a wind turbine actuated in surge motions. Surge amplitudes between $A = 0.150$ and 0.600 m ($A^* = A/D = 0.128$ and 0.514) were tested in combination with motion periods between $T = 0.5$ and 12 s. These combinations resulted in reduced frequencies $k = fD/u_1$ between 0.079 and 1.821 and nondimensional surge velocities $u^* = fA/u_1$ between 0.039 and 0.234. The specific combinations of $A$ and $T$ explored in this study, and their respective values of $u^*$, are given in Table \ref{tab:waveformparams}. Sinusoidal and trapezoidal surge-velocity waveforms served as motion profiles for these experiments. The trapezoidal waveforms consisted of alternating segments of constant acceleration and constant velocity. The relative duration of the constant-acceleration phases was parameterized by $\xi$, defined as the total time of nonzero acceleration in a single cycle divided by the cycle period. Hence, $\xi = 0$ corresponded to a square wave, while $\xi = 1$ corresponded to a triangle wave. The types of waveforms used in these experiments are shown in Figure \ref{fig:waveforms}.

\begin{table}
	\begin{center}
		\def~{\hphantom{0}}
		\begingroup
		\renewcommand{\arraystretch}{1.5} 
		\begin{tabular}{l||c|c|c|c}
			$u^*$ & $A^*=0.128$ & $A^*=0.257$ & $A^*=0.385$ & $A^*=0.514$ \\\hline\hline
			$T=12$ s & $-$ & $-$ & $-$ & 0.039 \\\hline
			$T=6$ s & $-$ & $-$ & $-$ & 0.078 \\\hline
			$T=3$ s & $-$ & $-$ & $-$ & 0.156 \\\hline
			$T=2$ s & $-$ & 0.117 & 0.175 & 0.234 \\\hline
			$T=1$ s & $-$ & 0.234 & $-$ & $-$ \\\hline
			$T=0.5$ s & 0.234 & $-$ & $-$ & $-$ \\
		\end{tabular}
		\endgroup
		\caption{Combinations of nondimensional amplitude $A^* = A/D$ and motion period $T$ tested in this study, tabulated with their respective nondimensional surge velocities $u^*$.}
		\label{tab:waveformparams}
	\end{center}
\end{table}

\begin{figure}
	\centering
	\includegraphics[width=\columnwidth]{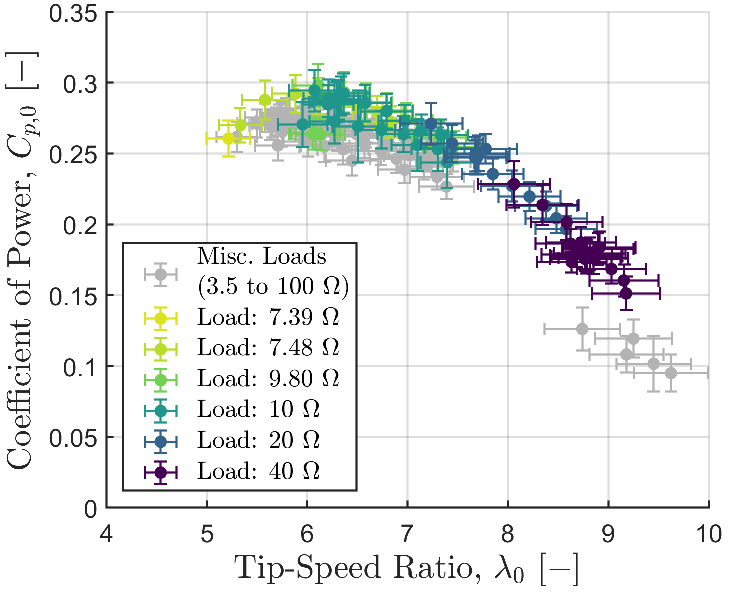}
	\caption{Steady power curve for the turbine used in this study, measured over a range of resistive loads (3.5 to 100 $\rm{\Omega}$) and wind speeds (6.05 to 12.09 $\rm{ms^{-1}}$).}
	\label{fig:CpTSR}
\end{figure}

The wind turbine was tested with resistive loads of 7.39, 7.48, 9.80, 10, 20, and 40 Ohms. The first three loads were attained using the rheostat and fixed resistors, and the second three were achieved using fixed 10-Ohm resistors in series. These corresponded to steady tip-speed ratios $\lambda_0 = R\omega_0/u_1$ between $5.21\pm0.22$ and $8.77\pm0.35$, and coefficients of power $C_{p,0} = \mathcal{P}_0/\frac{\pi}{2}\rho R^2 u_1^3$ between $0.176\pm0.010$ and $0.288\pm0.013$. A summary of the loading conditions and their corresponding steady nondimensional parameters is given in Table \ref{tab:ndparams}. Additionally, a power curve consisting of a collection of measurements at different wind speeds and loading conditions is shown in Figure \ref{fig:CpTSR}. Each of these steady measurements was conducted 3.09 m downstream of the fan array and over a duration of at least two minutes.

\begin{table*}
	\begin{center}
		\def~{\hphantom{0}}
		\begingroup
		\renewcommand{\arraystretch}{1.5} 
		\begin{tabular}{l|c|c|c|c|c|c}
			Resistive Load & 7.39 $\rm{\Omega}$ & 7.48 $\rm{\Omega}$ & 9.80 $\rm{\Omega}$ & 10 $\rm{\Omega}$ & 20 $\rm{\Omega}$ & 40 $\rm{\Omega}$ \\
			Case Identifier & 
			\includegraphics[width=10pt]{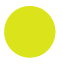} & \includegraphics[width=10pt]{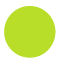} & \includegraphics[width=10pt]{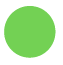} & \includegraphics[width=10pt]{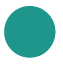} &
			\includegraphics[width=10pt]{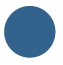} &
			\includegraphics[width=10pt]{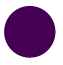} \\\hline\hline
			$\lambda_0$, sinusoidal cases & $5.21\pm0.22$ & $5.34\pm0.22$ & $6.11\pm0.25$ & $6.21\pm0.25$ & $7.72\pm0.31$ & $8.64\pm0.35$ \\
			$\lambda_0$, trapezoidal cases & $-$ & $-$ & $6.11\pm0.25$ & $6.27\pm0.26$ & $7.67\pm0.31$ & $8.77\pm0.35$ \\\hline
			$C_{p,0}$, sinusoidal cases & $0.261\pm0.013$ & $0.270\pm0.012$ & $0.264\pm0.011$ & $0.288\pm0.013$ & $0.250\pm0.012$ & $0.181\pm0.009$ \\
			$C_{p,0}$, trapezoidal cases & $-$ & $-$ & $0.264\pm0.011$ & $0.286\pm0.014$ & $0.249\pm0.012$ & $0.176\pm0.010$ \\\hline
			$K_\ell$ $\left[\rm{kg\frac{m}{s}}\right]$, sinusoidal cases & 0.458 & 0.456 & 0.445 & 0.444 & 0.422 & 0.409 \\
			$K_\ell$ $\left[\rm{kg\frac{m}{s}}\right]$, trapezoidal cases & $-$ & $-$ & 0.445 & 0.443 & 0.423 & 0.407 \\\hline
			$K_d$ $\left[\rm{kg\frac{m}{s}}\right]$, sinusoidal cases & 0.0262 & 0.0264 & 0.0276 & 0.0278 & 0.0302 & 0.0317 \\
			$K_d$ $\left[\rm{kg\frac{m}{s}}\right]$, trapezoidal cases & $-$ & $-$ & 0.0276 & 0.0279 & 0.0301 & 0.0319 \\\hline
			$K_1$ $\left[\rm{kg\frac{m^2}{s}}\right]$ & $-$ & 0.0141 & 0.0111 & 0.0112 & 0.00649 & 0.00376 \\\hline
			$K_0$ $\left[\rm{Nm}\right]$ & $-$ & 0.153 & 0.125 & 0.119 & 0.0850 & 0.0676 \\
		\end{tabular}
		\endgroup
		\caption{Performance characteristics and model constants for the six loading conditions investigated in this study. The values of $K_1$ and $K_0$ were not measured directly for the 7.39 $\rm{\Omega}$ case; the values from the 7.48 $\rm{\Omega}$ case were used instead.}
		\label{tab:ndparams}
	\end{center}
\end{table*}

\subsection{Experimental Procedure}
\label{sec:procedure}

Experiments were conducted in the CAST fan-array wind tunnel between May and July 2021. A zero-offset reading was taken for the torque sensor at the start of every day of measurements. To bring the turbine from rest to its prescribed operating condition, a higher wind speed was initially applied for at least two minutes to mitigate the effects of startup hysteresis in the shaft assembly. Each unsteady experiment was paired with a corresponding steady reference case, conducted within one hour of the unsteady case to minimize the influence of changing atmospheric conditions. The steady measurements were taken 3.09 m downstream of the fan array, defined as $x = 0$ (where $x$ is positive in the upstream direction). In the unsteady experiments, the turbine moved between $x = 0$ and $x = 2A$ at a prescribed frequency $f = 2\pi/T$. Each unsteady test began with five to ten startup periods to allow the system to reach cycle-to-cycle equilibrium. After these initial cycles, torque and rotation-rate measurements were recorded over 100 successive motion periods. For the shortest motion periods ($T = 0.5$ s), 200 motion periods were measured. The torque data were filtered and numerical derivatives of the angular-encoder readings were obtained, and these data were used to compute temporal means and time-resolved, phase-averaged profiles. The amplitudes and phases of these phase-averaged profiles were calculated by means of a fast Fourier transform. Finally, the aerodynamic torque was inferred via Equation \ref{eqn:swingEqn}, where $\tau_{gen}$ was supplied by the phase-averaged measured torque and $\frac{d\omega}{dt}$ was computed using a second-order central differencing scheme and was filtered using a sixth-order Butterworth filter with a cutoff frequency of 100 Hz to attenuate numerical-differentiation errors.

\subsection{Computing Model Constants}
\label{sec:constants}

To compare the experimental results with the model derived in Section \ref{sec:model}, empirical methods were developed to compute the coefficients of the model. Both the generator constants $K_1$ and $K_0$ and the aerodynamic model coefficients $K_\ell$ and $K_d$ were computed from steady torque measurements, without requiring any information from unsteady tests.

\begin{figure}[t]
	\centering
	\includegraphics[width=\columnwidth]{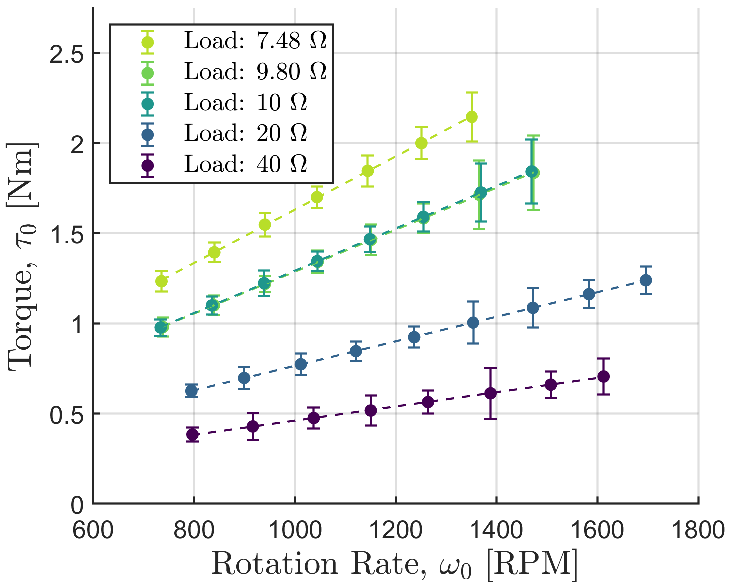}
	\caption{Measured torque from steady experiments with different resistive loads, plotted against rotor speed. The two coefficients from the linear fits (shown as dashed lines, $R^2 > 0.999$) correspond to the generator constants $K_1$ and $K_0$.}
	\label{fig:Tau_omega}
\end{figure}

The generator constants $K_1$ and $K_0$ were obtained by applying a constant resistive load to the turbine and measuring torque over a range of wind speeds. From these data, linear fits for the generator torque as a function of rotor speed could be extracted for each loading condition. The fit coefficients corresponded directly with the generator constants $K_1$ and $K_0$ from Equation \ref{eqn:tauGen}. The calculated values are reported in Table \ref{tab:ndparams} above, and, as expected for this type of generator, they scaled inversely with resistance. The accuracy of the generator-torque model depends primarily on the linearity of the generator within the typical operating conditions of the turbine. As evidenced by the data and linear fits shown in Figure \ref{fig:Tau_omega}, the generator used in this study fulfilled this condition well ($R^2>0.999$ in all cases). 

\begin{figure}[t]
	\centering
	\includegraphics[width=\columnwidth]{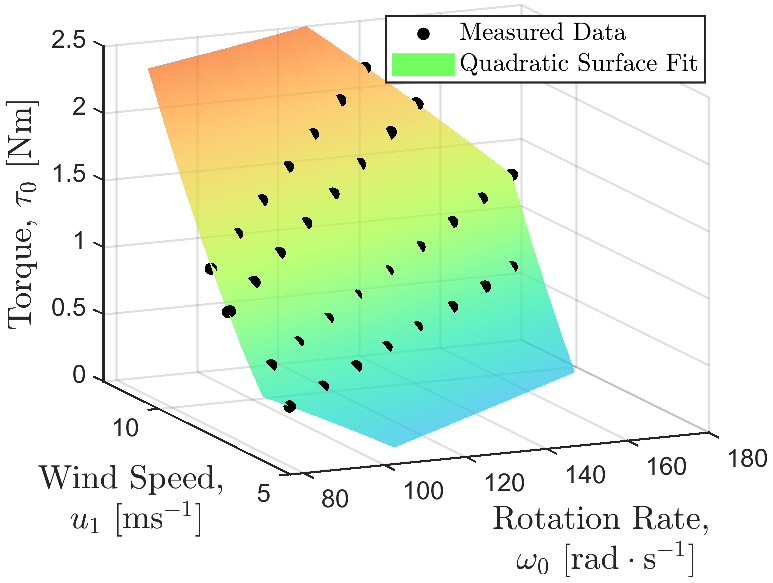}
	\caption{Measured torque from steady experiments with different resistive loads, plotted against both wind speed and rotor speed. A second-order polynomial surface ($R^2 > 0.999$) was fitted to the points to facilitate the computation of the linearized sensitivities $K_\ell$ and $K_d$.}
	\label{fig:Tau_surface}
\end{figure}

The same steady torque data were used to compute the aerodynamic model coefficients $K_\ell$ and $K_d$. The steady torque data was plotted as a three-dimensional set of points with respect to the wind speed and rotor speed, and a second-order polynomial surface fit was applied to the points. The fit equation gave an analytical expression for the partial derivatives with respect to $u_1$ and $\omega_0$ at any point, from which the coefficients $K_\ell$ and $K_d$ were calculated by means of the definitions given in Equations \ref{eqn:KL} and \ref{eqn:Kd}. The accuracy of the method as a quasi-steady aerodynamic representation is limited by the topology of the torque surface and the fidelity of the applied surface fit. For the turbine used in this study, the second-order polynomial surface (shown in Figure \ref{fig:Tau_surface}) was only weakly quadratic and fit the data with $R^2 > 0.999$, so the model was expected to perform well over the range of conditions tested in these experiments.

The moment of inertia of the generator was estimated from manufacturer specifications of the mass and radius of its rotor to be $K_2 = 6.96\times10^{-4}$ kg $\rm{m^2}$. The moment of inertia of the turbine and shaft assembly (including the generator), $J$, could be estimated in this manner as well. However, given the number of parts and nontrivial geometries in the assembly, a more empirical approach was taken. A 3D-printed spool with a diameter of 6 cm was attached to the turbine shaft, and the generator circuit was disconnected. A string was wrapped around the spool and connected to a weight, suspended from the turbine tower by means of a pulley. The weight was dropped ten times, and using the average measured torque and a fit based on the measured rotation-rate signal and the equation of motion of the system, an average moment of inertia was found to be $J = 0.0266\pm0.0008$ $\rm{kgm^2}$. This value aligned with the results of geometric estimates.

Thus, the only information required to obtain first-order predictions of the dynamics of a surging turbine were two moments of inertia and a series of steady torque measurements over a range of rotor speeds and wind speeds. Using the constants obtained from these prerequisites, the amplitude and phase of the aerodynamic and measured torques could be calculated analytically using the transfer functions given in Equations \ref{eqn:tfAero} and \ref{eqn:tfMeas}. To predict the time-resolved dynamics of the turbine for general surge-velocity waveforms, a numerical integration of the model as an initial-value problem was required. This was carried out using a fourth-order Runge-Kutta integration scheme with a time step of $10^{-3}$ s over ten motion periods, with the rotor speed initialized at $\omega\big\rvert_{t=0}=\omega_0$. The final period served as a representation of the steady-state model prediction.

\subsection{Sources of Uncertainty}
\label{sec:uncertainty}

Before the results of the experiments can be discussed in detail, the nature of the experimental facility requires a consideration of the sources of uncertainty in the measurements. These sources can be divided into two types: those that occurred over short time scales relative to a single test case, and those that occurred over longer time scales and were thus not captured in the error estimates computed from each data set.

Sources of uncertainty that occurred over short time scales contributed to the error bounds that will be shown in the figures in the following section. The standard deviation of the wind speed in the tunnel, measured over five minutes, was $2.20\pm0.17$\% of the average wind speed. This variability includes fluctuations due to turbulence and short-period fluctuations in the bulk flow velocity due to atmospheric disturbances. Though external winds were generally stronger during the afternoon, no corresponding increase in torque or rotation-rate uncertainty was evident for measurements conducted under these conditions. Therefore, despite the exposure of the facility to local atmospheric conditions, gusts and pressure fluctuations had a small influence on the results compared to other factors. Electrical noise from the torque sensor also contributed to measurement uncertainties, though these effects were reduced by the 100-Hz lowpass filter applied to the raw torque signal. Lastly, a slight misalignment in the turbine shaft was responsible for small variations in the measured torque within every full rotation of the turbine. These intracycle variations accounted for much of the uncertainty on the mean-torque measurements, and were occasionally visible in the phase-averaged torque and rotation-rate profiles. However, since their time scales were dictated by the rotor speed and were thus one to two orders of magnitude faster than the time scales of the surge oscillations, they did not directly affect the surge dynamics that were the focus of this study.

The open-air nature of the facility, however, meant that changing conditions throughout the day and across multiple days introduced additional sources of uncertainty that were not explicitly captured in the error estimates computed directly from each data set. Measurements of the wind speed at a single location over temperatures between 24.7 and $31.9^\circ$C showed a mild dependence of wind speed on temperature ($R^2=0.535$), resulting in a 3.0\% overall difference in wind speed. Given that temperatures in the facility ranged from 16.8 to $32.6^\circ$C across all experiments, a linear model would predict an uncertainty in the wind speed of $\pm3.2\%$. However, since no wind-speed measurements were taken at temperatures below $24^\circ$C, the fidelity of a linear model across the entire range of temperatures could not be directly confirmed. Because of these uncertainties, no temperature corrections were applied \textit{a posteriori} to the wind-speed data. This source of uncertainty complicated comparisons of mean torque and power between test cases, though this was to some extent ameliorated by normalizing the mean data from unsteady tests by a temporally proximal steady case. By contrast, the amplitude and phase depended primarily on the surge kinematics, which were very repeatable on account of the precision of the linear actuator, and were thus less affected by relatively small differences in wind speed.

In addition, the zero offset of the torque sensor exhibited a hysteretic dependence on temperature. To test this, the torque sensor was left in the facility for a period of 28 hours, and a voltage measurement was recorded every ten minutes. After the measurements were completed, during which the temperature in the facility ranged from 19.5 to $33.3^\circ$C, the difference between the initial and final voltage measurements corresponded to a torque difference of 0.014 Nm. Compared to the average torque measurements reported in this study, this represented a total shift of 1.2 to 2.9\%. Since the torque sensor was zeroed at the start of every day of experiments, this served as an upper bound on the uncertainty introduced by the zero-offset drift. This additional uncertainty again primarily obfuscated the mean measured torque, without affecting the amplitude and phase.

Finally, the intracycle torque variations described previously increased in magnitude in the tests conducted during June and July 2021, increasing the reported experimental error of the longest-period tests ($T = 6$ and 12 s). Higher temperatures and direct sunlight on the apparatus during these experiments may have caused the shaft assembly materials to expand, thereby exacerbating the rotational asymmetries in shaft friction responsible for the torque variations. This temperature dependence of friction could have influenced both the mean torque and power as well as their amplitude.

To demonstrate the cumulative effect of these sources of uncertainty, a series of 13 test cases, composed of six sinusoidal motion profiles at two different loading conditions, and a single sinusoidal motion profile at a third loading condition, was tested twice at different times of day. All of these data were included in the results presented in the following section for visual comparison. These thus serve as qualitative indicators of the influence of environmental conditions and system hysteresis on the results and analysis. As suggested previously, the effects of these factors will be most evident in the mean torque and power measurements, and to a lesser extent in the measured torque amplitudes.

\section{Experimental Results}
\label{sec:results}

Experiments were conducted with 32 different sinusoidal waveforms over 6 loading conditions and 42 different trapezoidal waveforms over 4 loading conditions and 4 values of the waveform parameter $\xi$. As mentioned previously, 13 of the sinusoidal cases were repeated at different times of day to convey the additional uncertainty due to changing conditions in the facility: 6 at $\lambda_0=6.11$, 6 at $\lambda_0=8.64$, and 1 at $\lambda_0=7.72$. The data from these experiments are presented in this section, in terms of torque amplitude, torque phase, and mean power. A selection of phase-averaged power measurements and their associated model predictions are provided as Figures \ref{fig:sine10Ohm} through \ref{fig:trap01} in Appendix \ref{sec:appPower}.


\subsection{Torque Amplitude Response}
\label{sec:results_amp}

\begin{figure*}[!ht]
	\begin{subfigure}[t]{\columnwidth}
		\centering
		\includegraphics[width=\textwidth]{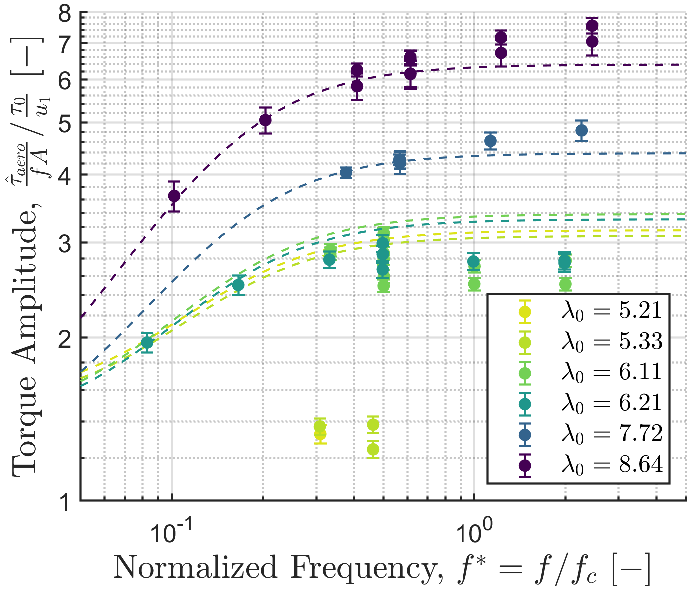}
		\caption{Sinusoidal surge waveforms.}
		\label{fig:ampAeroSine}
	\end{subfigure}
	\hfill
	\begin{subfigure}[t]{\columnwidth}
		\centering
		\includegraphics[width=\textwidth]{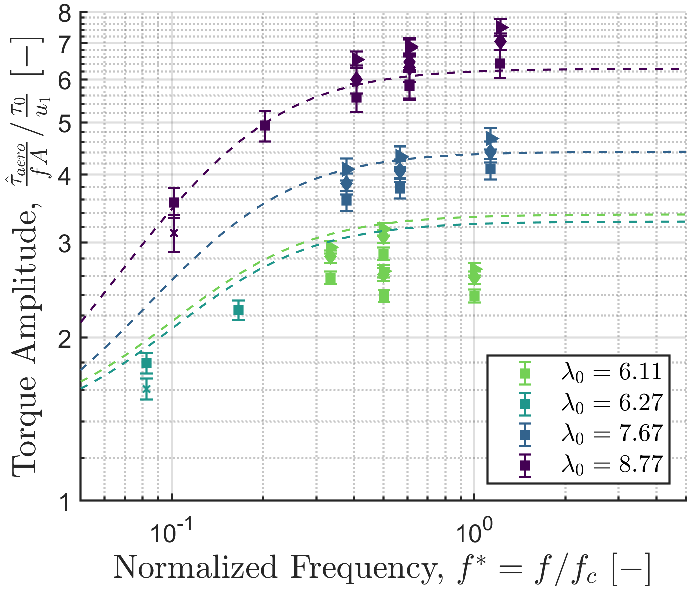}
		\caption{Trapezoidal surge waveforms. Markers represent $\xi = 0.01$ ($\times$), 0.25 ($\square$), 0.5 ($\diamond$), and 1 ($\triangleright$).}
		\label{fig:ampAeroTrap}
	\end{subfigure}
	\caption{Aerodynamic torque amplitude data (markers), compared with model predictions (dashed lines), for (a) sinusoidal and (b) trapezoidal surge-velocity waveforms.}
	\label{fig:ampAero}
\end{figure*}

\begin{figure*}[!ht]
	\begin{subfigure}[t]{\columnwidth}
		\centering
		\includegraphics[width=\textwidth]{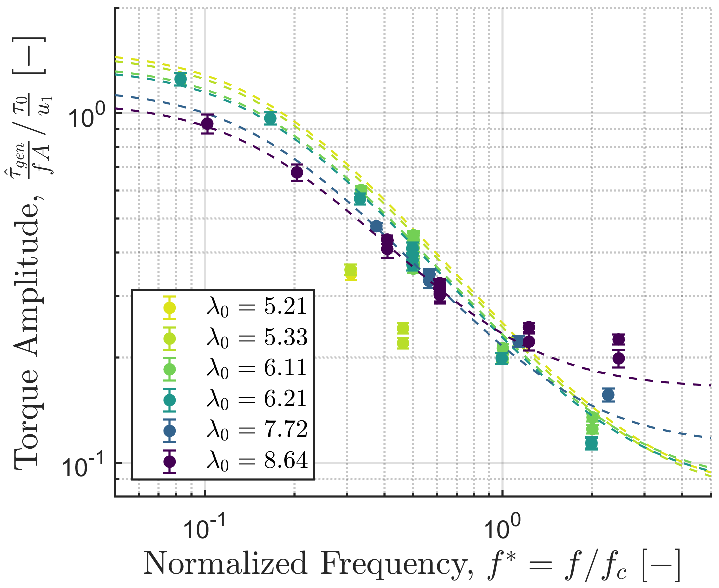}
		\caption{Sinusoidal surge waveforms.}
		\label{fig:ampMeasSine}
	\end{subfigure}
	\hfill
	\begin{subfigure}[t]{\columnwidth}
		\centering
		\includegraphics[width=\textwidth]{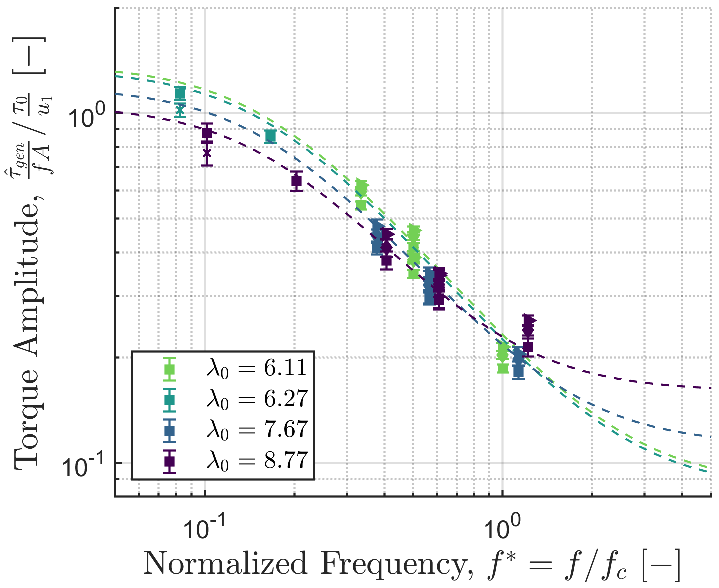}
		\caption{Trapezoidal surge waveforms. Markers represent $\xi = 0.01$ ($\times$), 0.25 ($\square$), 0.5 ($\diamond$), and 1 ($\triangleright$).}
		\label{fig:ampMeasTrap}
	\end{subfigure}
	\caption{Generator torque amplitude data (markers), compared with model predictions (dashed lines), for (a) sinusoidal and (b) trapezoidal surge-velocity waveforms.}
	\label{fig:ampMeas}
\end{figure*}

Aerodynamic-torque amplitudes, referenced to their respective surge-velocity amplitudes $U=fA$, are shown for sinusoidal and trapezoidal waveforms in Figure \ref{fig:ampAero}, while similar plots for the generator (measured) torque are given in Figure \ref{fig:ampMeas}. The transfer-function magnitudes were normalized by the steady torque for each case $\tau_0$ and the wind speed $u_1$. The aerodynamic torque was predicted by the model to increase in amplitude and asymptotically approach a maximum value above $f^*\approx1$. The generator torque predictions showed behavior more characteristic of a low-pass filter, where the amplitude decreased with increased frequency. The data for both the aerodynamic torque and generator torque showed good agreement with the model predictions for low values of $f^*$, except for the two lowest tip-speed ratios in the sinusoidal-waveform cases. Excluding these tip-speed ratios and the highest three values of $f^*$, the average relative error between the model predictions and measured data was 2.79\% and 2.13\% for the aerodynamic and generator torque amplitudes, respectively. At the two lowest tip-speed ratios, the turbine was close to stall, a flow regime not accounted for in the model. Decreasing the resistive load by 0.1 $\rm{\Omega}$ in the steady case caused the turbine to stall completely, slowing to a rotor speed of less than 10 $\rm{rads^{-1}}$. Unsteady tests undertaken at these two loading conditions with higher amplitudes and frequencies than those shown also caused the turbine to stall completely. Therefore, it was likely that the downstream surge motions caused the turbine to experience stall due to a reduction in the incident wind velocity relative to the turbine, thus decreasing the amplitude of the torque oscillations. A less severe decrease in amplitude could be seen at steady tip-speed ratios close to $\lambda_0\approx6.2$, which represented the optimal operating condition for the turbine (cf.\ Figure \ref{fig:CpTSR}). At normalized frequencies approaching $f^*=1$, the amplitude began to drop below the model prediction. Given the evidence of stall at lower tip-speed ratios, it is likely that this deviation was an effect of stall onset along a portion of the turbine blades as well. This conjecture is also in agreement with the simulations of \citeauthor{tran_cfd_2016}\cite{tran_cfd_2016}, who suggested that large surge motions at low to intermediate tip-speed ratios can cause stall to occur at the roots of the turbine blades and propagate radially outwards.

In contrast to the decreases in amplitude observed at low tip-speed ratios, the amplitudes at higher normalized frequencies ($f^*>1$) and tip-speed ratios above the optimal value increased above the model predictions. Increases in amplitude were also correlated with increasing $\xi$ (i.e.\ increasing proportions of streamwise acceleration within each motion cycle) in the trapezoidal-waveform experiments. The higher tip-speed ratios implied that the local angle of attack along the turbine blades was lower in these cases, reducing the likelihood that these increases in amplitude were due to stall phenomena. Since the model appeared to accurately capture the torque amplitudes at lower frequencies, it could be hypothesized that these trends were evidence of additional dynamics that became more salient at higher levels of unsteadiness. The specific nature of these dynamics cannot be identified definitively in the absence of flow measurements or a higher-order model for validation, but the present measurements do permit speculation as to the source of the observed discrepancies with the model. These considerations will be discussed in Section \ref{sec:discussion_discrepancies}.

\subsection{Torque Phase Response}
\label{sec:results_phase}

\begin{figure*}[!ht]
	\begin{subfigure}[t]{\columnwidth}
		\centering
		\includegraphics[width=\textwidth]{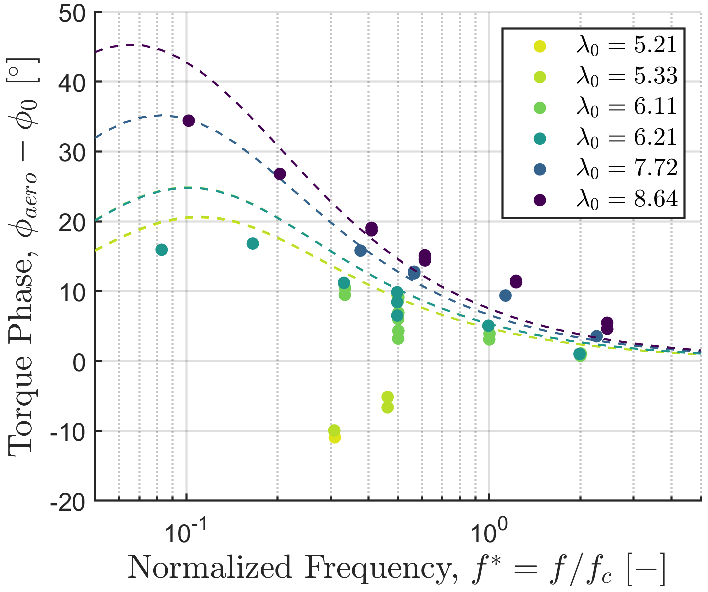}
		\caption{Sinusoidal surge waveforms.}
		\label{fig:phaseAeroSine}
	\end{subfigure}
	\hfill
	\begin{subfigure}[t]{\columnwidth}
		\centering
		\includegraphics[width=\textwidth]{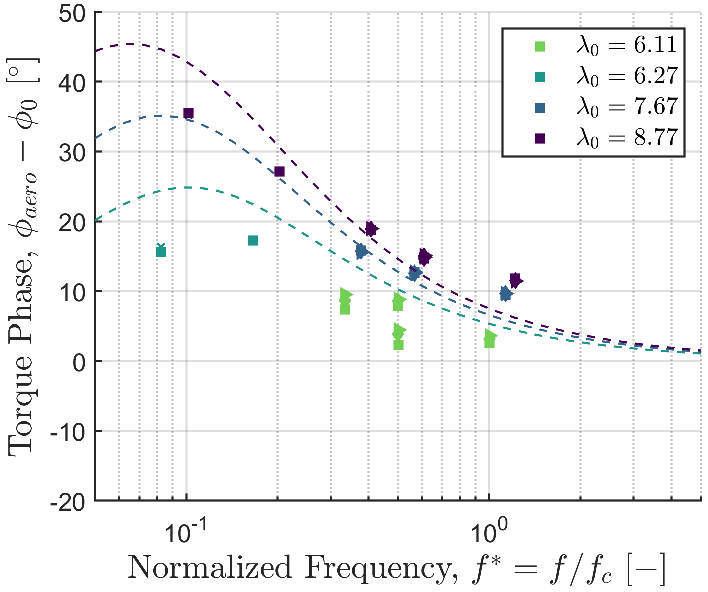}
		\caption{Trapezoidal surge waveforms. Markers represent $\xi = 0.01$ ($\times$), 0.25 ($\square$), 0.5 ($\diamond$), and 1 ($\triangleright$).}
		\label{fig:phaseAeroTrap}
	\end{subfigure}
	\caption{Aerodynamic torque phase data (markers), compared with model predictions (dashed lines), for (a) sinusoidal and (b) trapezoidal surge-velocity waveforms.}
	\label{fig:phaseAero}
\end{figure*}

\begin{figure*}[!ht]
	\begin{subfigure}[t]{\columnwidth}
		\centering
		\includegraphics[width=\textwidth]{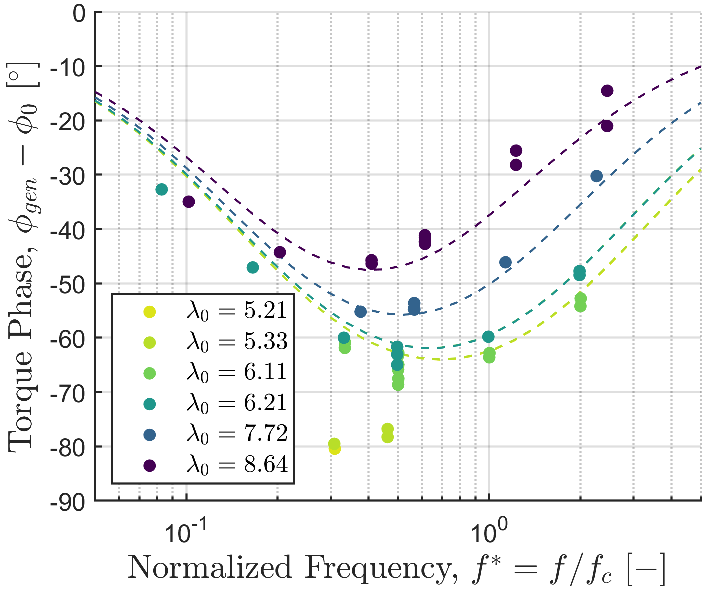}
		\caption{Sinusoidal surge waveforms.}
		\label{fig:phaseMeasSine}
	\end{subfigure}
	\hfill
	\begin{subfigure}[t]{\columnwidth}
		\centering
		\includegraphics[width=\textwidth]{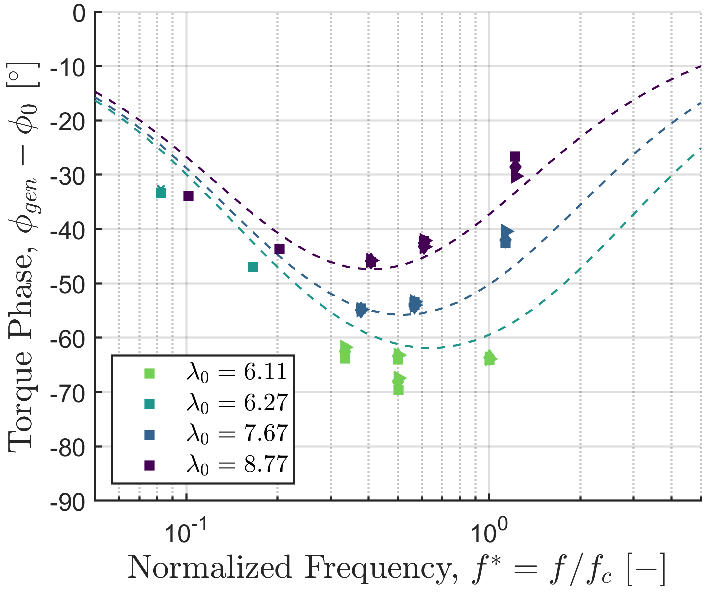}
		\caption{Trapezoidal surge waveforms. Markers represent $\xi = 0.01$ ($\times$), 0.25 ($\square$), 0.5 ($\diamond$), and 1 ($\triangleright$).}
		\label{fig:phaseMeasTrap}
	\end{subfigure}
	\caption{Generator torque phase data (markers), compared with model predictions (dashed lines), for (a) sinusoidal and (b) trapezoidal surge-velocity waveforms.}
	\label{fig:phaseMeas}
\end{figure*}

The aerodynamic and generator phase-response data for the same experimental cases shown in Figures \ref{fig:ampAero} and \ref{fig:ampMeas} are given in Figures \ref{fig:phaseAero} and \ref{fig:phaseMeas} and are compared with model predictions. The generator torque phase lagged behind the forcing signal $U(t)$ across all tested frequencies, while the aerodynamic torque phase led the forcing signal for all cases except those at the lowest two tip-speed ratios. Excluding the lowest two tip-speed ratios, the average difference between the model predictions and measured data was $3.59^\circ$ and $3.21^\circ$ for the aerodynamic and generator torque phase, respectively. The phase in the anomalous cases showed an approximately $20^\circ$ lag relative to the predictions of the model, suggesting again that the turbine blades were experiencing the effects of stall under these conditions. The tests conducted at higher tip-speed ratios followed the qualitative trends predicted by the model, though the model slightly overpredicted the aerodynamic and generator torque phase by around $6^\circ$ at low normalized frequencies. Unlike the torque amplitude, the torque phase was relatively insensitive to changes in the trapezoidal-waveform parameter $\xi$.

\subsection{Mean Power}
\label{sec:results_power}

\begin{figure*}[!ht]
	\begin{subfigure}[t]{\columnwidth}
		\centering
		\includegraphics[width=\textwidth]{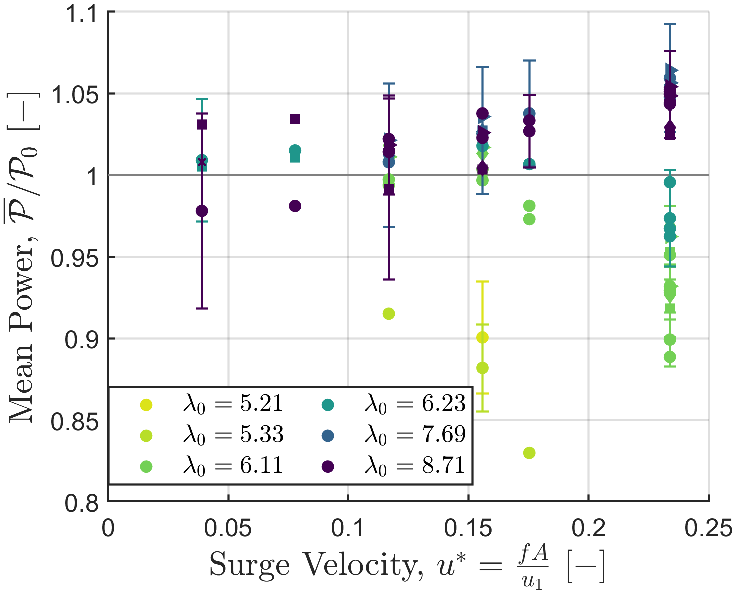}
		\caption{}
		\label{fig:P_U}
	\end{subfigure}
	\hfill
	\begin{subfigure}[t]{\columnwidth}
		\centering
		\includegraphics[width=\textwidth]{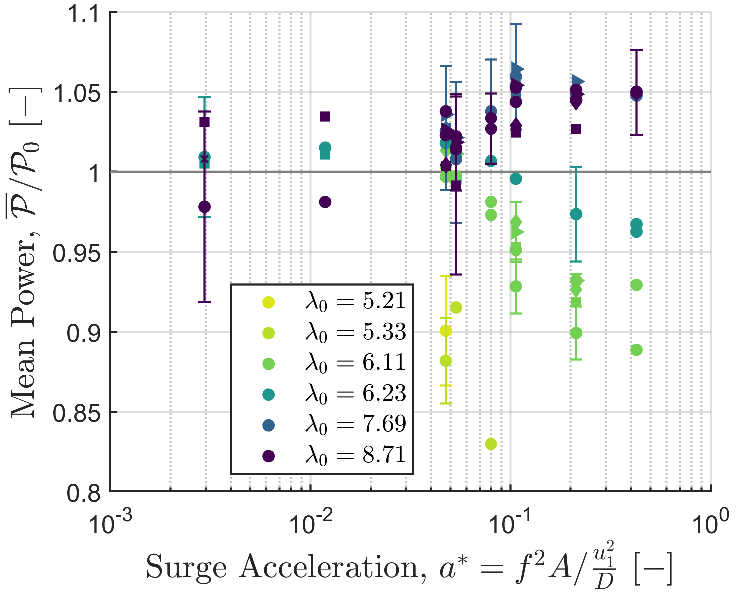}
		\caption{}
		\label{fig:P_a}
	\end{subfigure}
	\caption{Period-averaged measured power $\overline{\mathcal{P}}$, normalized by the reference steady power $\mathcal{P}_0$, plotted for all cases against (a) the nondimensional surge velocity $u^*$ and (b) the nondimensional surge acceleration $a^*$. Circular markers ($\circ$) represent sinusoidal waveform cases; non-circular markers represent trapezoidal-waveform cases with $\xi = 0.01$ ($\times$), 0.25 ($\square$), 0.5 ($\diamond$), and 1 ($\triangleright$). For the sake of clarity, error bars are only plotted for every sixth point.}
	\label{fig:meanP}
\end{figure*}

While the amplitude and phase responses of the aerodynamic and generator torque were nontrivial, the first-order linear model predicted zero change in the mean torque and power for all surge motions. The period-averaged measured torque and power, however, diverged from these predictions as the surge motions became more pronounced. When plotted against the nondimensional surge velocity $u^*$ and a nondimensional surge acceleration defined as $a^* = f^2 A / \frac{u_1^2}{D}$ (cf.\ Figure \ref{fig:meanP}), the normalized period-averaged measured power $\overline{\mathcal{P}}/\mathcal{P}_0$ diverged from unity at higher levels of unsteadiness. For tip-speed ratios at or below the optimal operating point, the average power decreased with increasing surge velocity and acceleration. This was especially the case for the lowest two tip-speed ratios tested; at higher surge velocities than those shown, the turbine stalled completely and the normalized average power dropped nearly to zero. It is thus reasonable to infer that the occurrence of stall along portions of the turbine blades was responsible for the decrease in power observed at lower tip-speed ratios and higher surge velocities. 

For higher tip-speed ratios, the average power increased with increasing surge velocity and acceleration, with a maximum measured value of $\overline{\mathcal{P}}/\mathcal{P}_0 = 1.064\pm0.045$. The increases appeared well-correlated with these surge parameters, whereas the specific type of waveform did not appear to significantly affect the results. Furthermore, the apparent consistency in trend over a range of testing conditions and times of day suggests that these results cannot simply be attributed to long-period variations in the conditions in the facility. These trends were similar to those shown by \citeauthor{farrugia_study_2016}\cite{farrugia_study_2016} (cf.\ Fig.\ 8 in their paper) and \citeauthor{wen_influences_2017}\cite{wen_influences_2017} (cf.\ Fig.\ 15b in their paper), though those studies focused on the reduced frequency as the unsteady parameter of interest. These large deviations from the predictions of the first-order linear model will be discussed in the following section.

\section{Discussion}
\label{sec:discussion}

\subsection{Performance of the Model}
\label{sec:discussion_model}

As shown in Sections \ref{sec:results_amp} and \ref{sec:results_phase}, as well as in the results presented in Appendix \ref{sec:appPower}, the first-order linear model was able to capture trends in amplitude and phase for torque and power, using only \textit{a priori} measurements from the turbine under steady-flow conditions. The time-resolved model predictions for power were qualitatively similar to the measured phase-averaged data shown in Figures \ref{fig:sine10Ohm} through \ref{fig:trap01} for all surge-velocity waveform types, both sinusoidal and trapezoidal. The model was also able to account for the effects of changes in the aerodynamic and generator parameters of the system, particularly through the characteristic frequency $f_c$. These results demonstrated that while the aerodynamic torque tends to increase in amplitude with frequency up to $f^*\approx1$ and leads the input surge-velocity waveform, the generator acts as a low-pass filter on the measured torque and power, leading to decreases in amplitude and phase lags.

The nontrivial phase prediction is particularly noteworthy, since previous studies showed little consensus in the phases of their torque and power data. The existing quasi-steady models in the literature (e.g.\ \citeauthor{mancini_characterization_2020}\cite{mancini_characterization_2020},  \citeauthor{johlas_floating_2021}\cite{johlas_floating_2021}) predicted zero phase difference with the input surge-velocity waveform. The present results have demonstrated that the aerodynamic and generator characteristics of the system produce phase differences of up to $60^\circ$, with greater differences observed in cases at low tip-speed ratios where stall is purported to have an effect. The success of the first-order linear model in recovering these trends over a wide range of loading conditions, surge frequencies, and surge amplitudes suggests that the model can help explain the variation in phase trends reported in the literature, and serve as an analytical foundation for future studies involving the time-resolved dynamics and control of surging turbines.

\subsection{Discrepancies Between Measurements and Model Predictions}
\label{sec:discussion_discrepancies}

Though the model showed good agreement with the measured data in terms of amplitude and phase, some discrepancies were still evident. The model tended to overpredict the torque amplitude at low tip-speed ratios and high surge frequencies and underpredict the torque amplitude at high tip-speed ratios and high surge frequencies. The model also did not correctly predict the trends in period-averaged power at high surge velocities. These deviations and their implications merit further discussion.

The observed decreases in torque amplitude and phase, as well as period-averaged power, were attributed to the onset of stall on the turbine blades. The turbine was close to static stall at the two lowest tip-speed ratios tested, as any slight decrease in the resistive load or inflow velocity would cause its rotor speed to fall to nearly zero. This observed behavior is in accordance with the assumption that the turbine blades were designed to maximize their lift-to-drag ratio near the optimal operating condition of $\lambda_0\approx6.2$, as lower tip-speed ratios would imply higher induced angles of attack that would bring the blade sections closer to the static stall angle. If this was the case, then at tip-speed ratios near or below the optimal operating condition of the turbine, increased induced angles of attack on the blades through the addition of unsteady surge motions could be expected to increase the local angle of attack above the static stall angle. This explanation aligns well with the observed decreases in the amplitude, phase and time-averaged responses of the system with increasing normalized frequency and decreasing tip-speed ratio. It would be useful to verify this conjecture with flow measurements or numerical simulations in the future.

The discrepancies in which the model underpredicts the measured data, by contrast, have physical sources that are more difficult to identify given the present data. For the two highest tip-speed ratios tested, the torque amplitude data exceeded the model predictions at higher normalized frequencies. Furthermore, in the trapezoidal-waveform cases shown in Figure \ref{fig:ampAeroTrap}, the torque amplitudes for cases with identical surge frequencies and amplitudes consistently increased with increasing $\xi$ (i.e.\ an increasing proportion of surge acceleration within each motion cycle). These increases in amplitude run counter to the predictions of classical unsteady-aerodynamic theory for an isolated, two-dimensional airfoil. The Sears function, an analytical transfer function for the unsteady lift response of a two-dimensional airfoil in a transverse gust, predicts a decrease in the lift amplitude relative to the quasi-steady case for a turbine blade section undergoing surge motions.\cite{sears_aspects_1941} A reduction in torque amplitude with increasing reduced frequency would thus be expected if the Sears function is a sufficient model for the unsteady aerodynamics of a surging turbine. Since the opposite trend is observed in the data, additional dynamics must be involved to account for the divergence from the model predictions, such as spanwise flow along the blades, blade-wake interactions, or other unsteady effects.

The divergence in the period-averaged power data from the model further underscores the need to appeal to additional dynamics. The linear nature of the model dictated that the period-averaged power for a periodically surging turbine would never differ from the steady reference power, irrespective of surge kinematics. Thus, the increases in the period-averaged power over the steady reference case observed at higher levels of unsteadiness must have been the result of effects that were not accounted for in the linear model. 

Either the assumption of linear aerodynamics or the assumption of quasi-steady aerodynamics could be called into question to explain the observed enhancements in torque amplitude and period-averaged power. For instance, the quasi-steady model of \citeauthor{johlas_floating_2021}\cite{johlas_floating_2021}, which is derived from the scaling of power with the cube of the incident wind velocity and is thus nonlinear, predicts that the period-averaged power should scale with $\left(u^*\right)^2$. This trend qualitatively agrees with the average-power data shown in Figure \ref{fig:P_U}, though their model does not capture the time-resolved turbine dynamics. This observation suggests that an incorporation of the nonlinear dependence of power on $u^3$ into the first-order linear model could better account for the improvements in period-averaged power with increasing $u^*$.

\begin{figure*}[!ht]
	\begin{subfigure}[t]{\columnwidth}
		\centering
		\includegraphics[width=\textwidth]{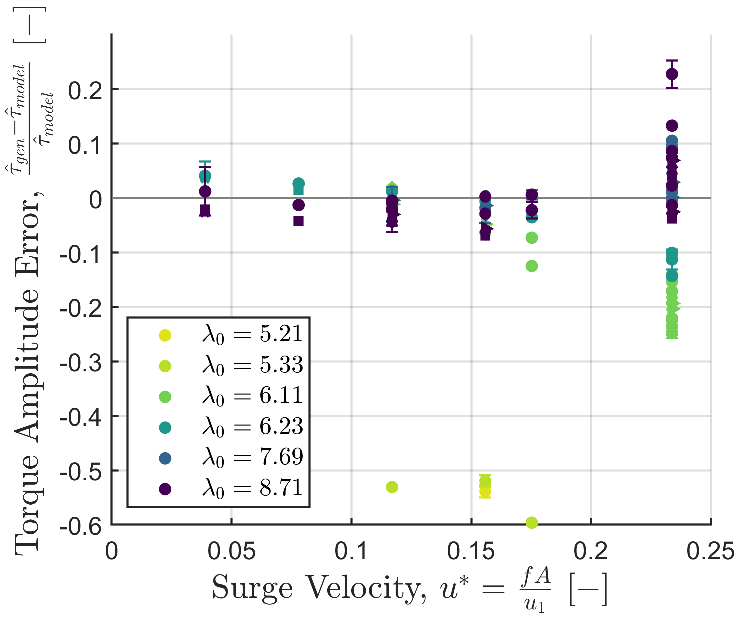}
		\caption{}
		\label{fig:ampErrU}
	\end{subfigure}
	\hfill
	\begin{subfigure}[t]{\columnwidth}
		\centering
		\includegraphics[width=\textwidth]{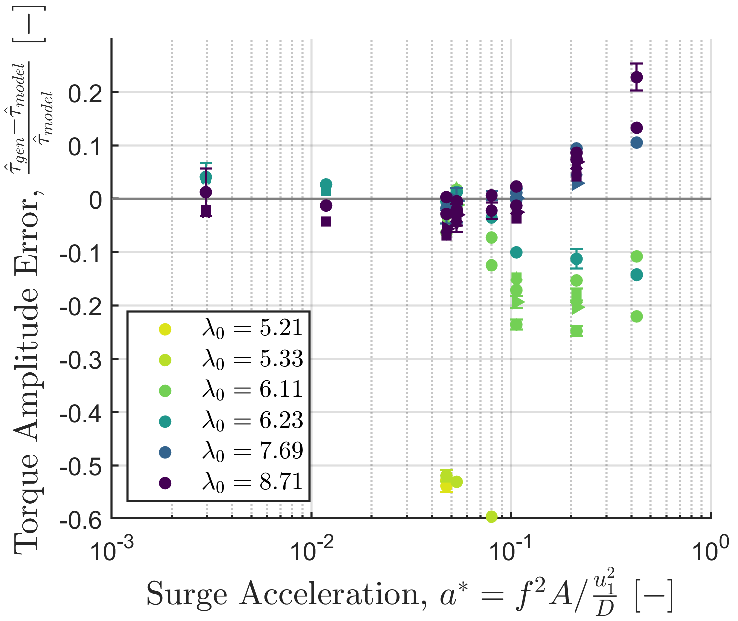}
		\caption{}
		\label{fig:ampErra}
	\end{subfigure}
	\caption{Relative error between measured and modelled torque amplitudes, plotted for all cases against (a) the nondimensional surge velocity $u^*$ and (b) the nondimensional surge acceleration $a^*$. Circular markers ($\circ$) represent sinusoidal waveform cases; non-circular markers represent trapezoidal-waveform cases with $\xi = 0.01$ ($\times$), 0.25 ($\square$), 0.5 ($\diamond$), and 1 ($\triangleright$). For the sake of clarity, error bars are only plotted for every sixth point.}
	\label{fig:ampErr}
\end{figure*}

Alternatively, the enhancements in torque amplitude and average power in the unsteady cases at high tip-speed ratios could be the result of unsteady fluid mechanics. As a proxy for flow unsteadiness, we briefly consider the surge acceleration. Returning to the apparently systematic increases in torque amplitude with increasing $\xi$, we recall that higher values of the parameter $\xi$ corresponded to waveforms in which segments of constant acceleration occupied a larger fraction of each period (with $\xi=1$ representing one half-cycle of constant acceleration, followed by one half-cycle of constant deceleration). Therefore, the increase in torque amplitude with $\xi$ could be interpreted as an effect of surge acceleration. To extend this point further, the relative error between the measured and modelled torque amplitude was plotted with respect to the nondimensional surge velocity $u^*$ and surge acceleration $a^*$ in Figure \ref{fig:ampErr}. While the data in Figure \ref{fig:ampErrU} became multi-valued at the highest value of $u^*$, more uniform trends were evident when the data were plotted against $a^*$ in Figure \ref{fig:ampErra}, suggesting that the surge acceleration was a more robust indicator of the difference between measured and predicted torque amplitudes. Finally, the apparent lack of scatter in the period-averaged power data when plotted against $a^*$ might similarly suggest that the surge acceleration could play a systematic role in producing the increases in average power, though the strength of this argument is limited by the fact that the definition of $a^*$ involves the same kinematic parameters as that of $u^*$. The role of surge acceleration cannot necessarily be accounted for in a linear sense, since adding a linear acceleration-dependent term to the existing model would introduce a $90^\circ$ phase lead that would disrupt the accuracy of the model with respect to phase (cf.\ Figures \ref{fig:phaseAero} and \ref{fig:phaseMeas}). Still, these observations do suggest that the discrepancies with respect to the first-order linear model are systematic, and that the physics behind these systematic deviations may include unsteady effects.

Flow-field measurements at both the blade and rotor scale would greatly facilitate the identification of unsteady contributions to the turbine dynamics. \citeauthor{el_makdah_influence_2019}\cite{el_makdah_influence_2019} provided some evidence of unsteady effects in their investigations of the wake structure of an accelerating low-inertia rotor, which were conducted in a water channel using planar particle-image velocimetry. A similar study at higher Reynolds numbers and involving periodic rather than unidirectional surge motions, and with the first-order linear model derived in this study as a baseline for torque and power comparisons, would be very informative for establishing the relative influence of nonlinear and unsteady effects. Without such quantitative flow measurements, however, it is difficult to definitively conclude at this stage that either quasi-steady nonlinear effects or purely unsteady aerodynamic effects are responsible for the observed enhancements in torque amplitude and period-averaged power. The main conclusion to be drawn from these speculative considerations is that enhancements in torque and power due to unsteady streamwise motions are possible, and that these scale systematically with the surge kinematics. If unsteady fluid mechanics do play a role in these enhancements, as \citeauthor{dabiri_theoretical_2020}\cite{dabiri_theoretical_2020} has suggested analytically, then additional experimental and modelling studies could uncover novel solutions that leverage these unsteady effects to produce higher period-averaged turbine efficiencies.

\subsection{Applications to Full-Scale Wind Turbines}
\label{sec:discussion_extensions}

Though these experiments were conducted at diameter-based Reynolds numbers that were one to two orders of magnitude smaller than those encountered in modern full-scale wind turbines, the dual analytical-empirical nature of the first-order linear model should in principle allow for a seamless extension to operational systems in field conditions. The topology of the torque space spanned by $u_1$ and $\omega$ will be different for full-scale systems, given the differences in full-scale turbine design and aerodynamics. A lack of Reynolds-number invariance in these results may also play a role in the precise topology of the torque space, since the chord-based Reynolds numbers encountered in these experiments ($5.6\times10^4 \lesssim Re_c \lesssim 9.2\times10^4$) were much lower than the invariance criterion of $Re_c \geq 3.5\times10^6$ proposed by \citeauthor{miller_horizontal_2019}\cite{miller_horizontal_2019}. Still, as long as a sufficiently reliable local linearization is possible, the procedure outlined in Section \ref{sec:constants} for computing the aerodynamic constants $K_\ell$ and $K_d$ should continue to give accurate predictions of time-resolved dynamics. Methods for characterizing the generator constants $K_0$, $K_1$, and $K_2$ may differ for full-scale generators, and gearbox effects may need to be taken into account, but these do not represent fundamental challenges to the validity of the model at scale. Thus, while to our knowledge no experiments with surging full-scale wind turbines have been carried out, the first-order linear model derived in this work should allow the dynamics of full-scale systems to be predicted solely on the basis of information from steady measurements, which may be readily obtained from onshore or fixed-bottom offshore turbines.

For floating offshore wind turbines, the surge amplitudes and velocities experienced due to typical surface-wave forcing patterns will likely be much smaller than the largest motions investigated in this study. Still, even for small perturbations, the disambiguation between aerodynamic and generator torque that the first-order linear model provides could inform design considerations and control strategies that leverage generator and turbine inertia or dynamic load-control schemes to reduce surge-induced fatigue loads on the turbine blades. These strategies could also be used to mitigate the effects of blade stall that may be encountered even in small surge motions if the turbine blades are operating at high static angles of attack.

In the extreme cases in which large surge excursions do occur, the surge motions of a FOWT would be coupled with large tilt disturbances, an effect not covered in this study. Therefore, it is questionable whether a FOWT will experience the large pure-surge oscillations represented by the maximally unsteady cases in these experiments. However, these kinds of surge kinematics may be possible in more nascent wind-energy technologies, such as kite-based airborne wind turbines. In these contexts, the large surge motions could be leveraged to increase the efficiency of the turbines above their steady values. The first-order linear model would inform decisions regarding trade-offs between surge-induced efficiency gains and the onset of stall at high surge-velocity amplitudes, as well as the aerodynamic design of the turbine blades themselves for these inherently unsteady environments.

Finally, the potential of the time-averaged power enhancement empirically demonstrated in this study should not be missed. The increasing size of wind turbines and wind farms, combined with logistical difficulties involving siting for new wind-power plants, makes traditional wind power increasingly challenging to implement on a global scale. The theoretical limitation of the efficiency of wind-energy devices to 59.3\%\cite{betz_maximum_1920} also remains a fundamental roadblock to the engineering and economic efficacy of wind power. Thus, if the overall capacity of wind power is to be expanded over the next few decades, then innovative solutions for increasing the efficiency of wind-energy systems must be considered. The land-independent nature of floating offshore wind farms and airborne wind-energy systems already represents an appealing solution to the problem of land-based siting. If the unsteady surge motions inherent to these systems and other nascent technologies could be leveraged for increased efficiency, the constraints on both efficiency and large-scale implementation could potentially be circumvented. 

\section{Conclusions}
\label{sec:conclusions}

In this study, the torque, rotor speed, and power of a horizontal-axis wind turbine undergoing periodic surge motions were investigated. A first-order linear model was derived to explain trends in amplitude and phase, and the experimental results compared favorably with the predictions of the model. Deviations from the model predictions were observed at low tip-speed ratios, a behavior that was attributed to the onset of stall on the turbine blades. At high tip-speed ratios, enhancements in the torque amplitude and period-averaged power were observed at high normalized surge frequencies. While the relative contributions of unsteady and nonlinear effects to these enhancements cannot be separated based on these experiments, it is nonetheless noteworthy that periodic surge motions can lead to increased period-averaged turbine efficiencies relative to the steady case. The trends captured by the model and data are expected to hold qualitatively for utility-scale wind turbines, both in the floating-offshore context and in novel applications that inherently leverage unsteady flow physics for increased power-conversion efficiencies.

\section*{Acknowledgements}

The authors gratefully acknowledge the assistance of several people, without whom the construction and operation of the experimental apparatus would not have been possible: J.\ Benson, who graciously provided machine-shop access during the pandemic; G.\ Juarez and M.\ Vega, who installed the power systems for the linear actuator; M.\ Miller, K.\ Bankford, and J.\ Kissing for technical consultation regarding the turbine power-control system and linear actuator; A.\ Kiani and E.\ Tang for machining key components of the apparatus; N.\ Esparza-Duran and R.\ Nemovi, who oversaw operations at CAST; M.\ Veismann and P.\ Renn for wind-tunnel support; and J.\ Cardona, E.\ Tang, P.\ Gunnarson, M.\ Fu, and R.\ Goldschmid for providing assistance and safety supervision for the experiments.

This work was funded by the National Science Foundation (grant CBET-2038071) and the Caltech Center for Autonomous Systems and Technologies. N.\ Wei was supported by a Stanford Graduate Fellowship and a National Science Foundation Graduate Research Fellowship.

\section*{Author Declarations}

The authors have no conflicts of interest to disclose. 

\section*{Data Availability}

The data that support the findings of this study are available from the corresponding author upon reasonable request.

\appendix

\section{Phase-Averaged Power Profiles}
\label{sec:appPower}

The plots shown in this section compare the measured, phase-averaged power over a single period (blue) with the prediction of the first-order linear model (orange) and the steady reference power (green). The model prediction was computed using a fourth-order Runge-Kutta time-integration scheme with a time-step of $10^{-3}$ s. Ten periods were simulated to attenuate startup effects, and the last period was extracted to represent the model prediction. The model did not capture the time-averaged power for the unsteady cases, but good agreement was still observed in terms of amplitude, phase, and waveform shape.

\begin{figure*}
	\begin{subfigure}[t]{\columnwidth}
		\centering
		\includegraphics[width=\textwidth]{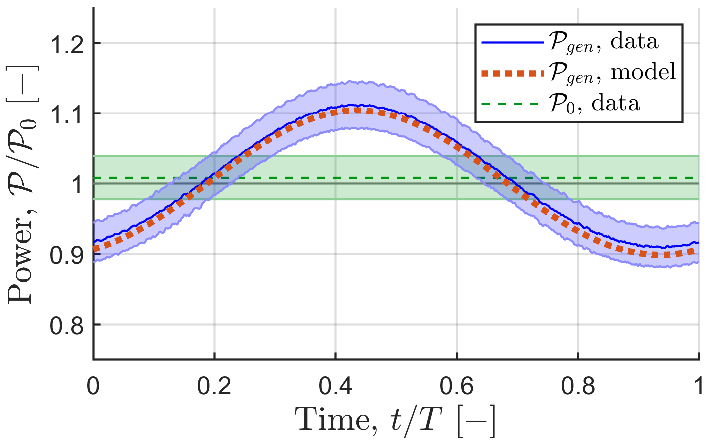}
		\caption{$A^* = 0.257$, $k = 0.455$, $u^* = 0.117$}
		\label{fig:sine10Ohm1}
	\end{subfigure}
	\hfill
	\begin{subfigure}[t]{\columnwidth}
		\centering
		\includegraphics[width=\textwidth]{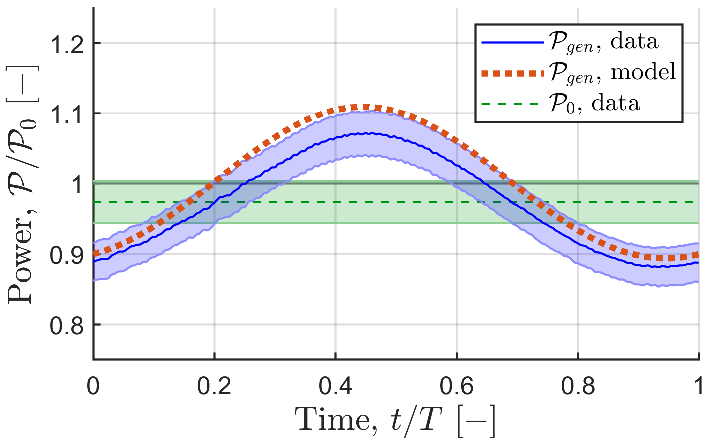}
		\caption{$A^* = 0.257$, $k = 0.911$, $u^* = 0.234$}
		\label{fig:sine10Ohm2}
	\end{subfigure}
	\begin{subfigure}[t]{\columnwidth}
		\centering
		\includegraphics[width=\textwidth]{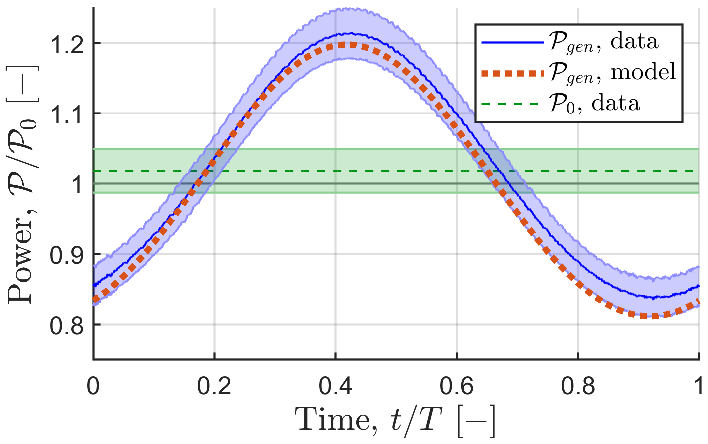}
		\caption{$A^* = 0.514$, $k = 0.304$, $u^* = 0.156$}
		\label{fig:sine10Ohm3}
	\end{subfigure}
	\hfill
	\begin{subfigure}[t]{\columnwidth}
		\centering
		\includegraphics[width=\textwidth]{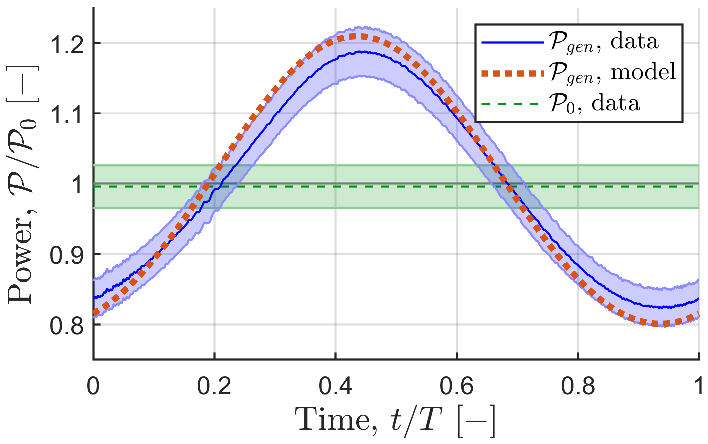}
		\caption{$A^* = 0.514$, $k = 0.455$, $u^* = 0.234$}
		\label{fig:sine10Ohm4}
	\end{subfigure}
	\caption{Sinusoidal waveforms with a load of 10 $\rm{\Omega}$ ($\lambda_0 = 6.21\pm0.25$), for four representative cases: (a) $A = 0.3$ m and $T = 2$ s, (b) $A = 0.3$ m and $T = 1$ s, (c) $A = 0.6$ m and $T = 3$ s, and (d) $A = 0.6$ m and $T = 2$ s.}
	\label{fig:sine10Ohm}
\end{figure*}

\begin{figure*}
	\begin{subfigure}[t]{\columnwidth}
		\centering
		\includegraphics[width=\textwidth]{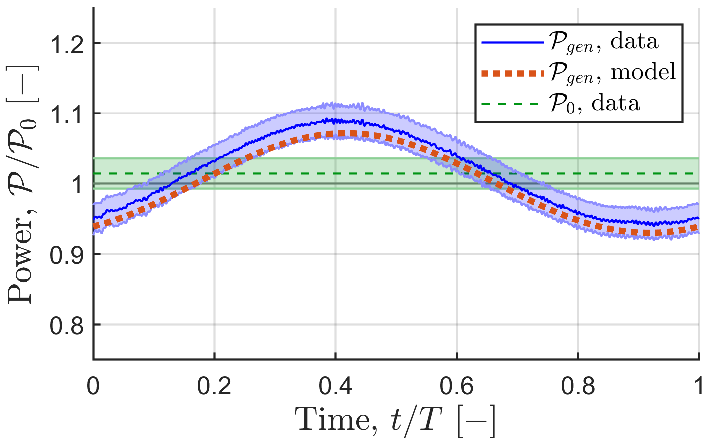}
		\caption{$A^* = 0.257$, $k = 0.455$, $u^* = 0.117$}
		\label{fig:sine40Ohm1}
	\end{subfigure}
	\hfill
	\begin{subfigure}[t]{\columnwidth}
		\centering
		\includegraphics[width=\textwidth]{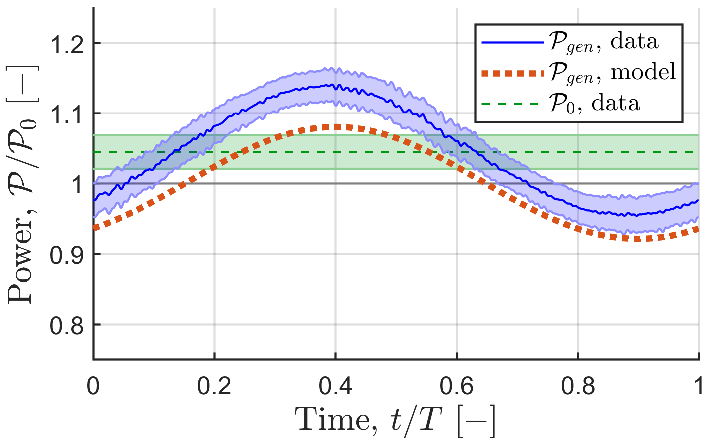}
		\caption{$A^* = 0.257$, $k = 0.911$, $u^* = 0.234$}
		\label{fig:sine40Ohm2}
	\end{subfigure}
	\begin{subfigure}[t]{\columnwidth}
		\centering
		\includegraphics[width=\textwidth]{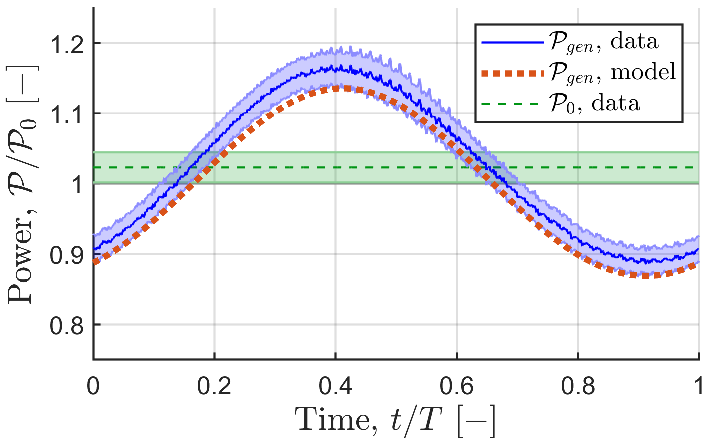}
		\caption{$A^* = 0.514$, $k = 0.304$, $u^* = 0.156$}
		\label{fig:sine40Ohm3}
	\end{subfigure}
	\hfill
	\begin{subfigure}[t]{\columnwidth}
		\centering
		\includegraphics[width=\textwidth]{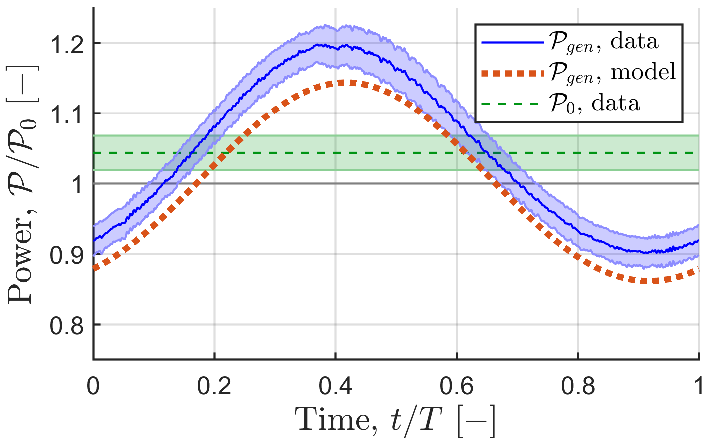}
		\caption{$A^* = 0.514$, $k = 0.455$, $u^* = 0.234$}
		\label{fig:sine40Ohm4}
	\end{subfigure}
	\caption{Sinusoidal waveforms with a load of 40 $\rm{\Omega}$ ($\lambda_0 = 8.64\pm0.35$), for four representative cases: (a) $A = 0.3$ m and $T = 2$ s, (b) $A = 0.3$ m and $T = 1$ s, (c) $A = 0.6$ m and $T = 3$ s, and (d) $A = 0.6$ m and $T = 2$ s.}
	\label{fig:sine40Ohm}
\end{figure*}

\begin{figure*}
	\begin{subfigure}[t]{\columnwidth}
		\centering
		\includegraphics[width=\textwidth]{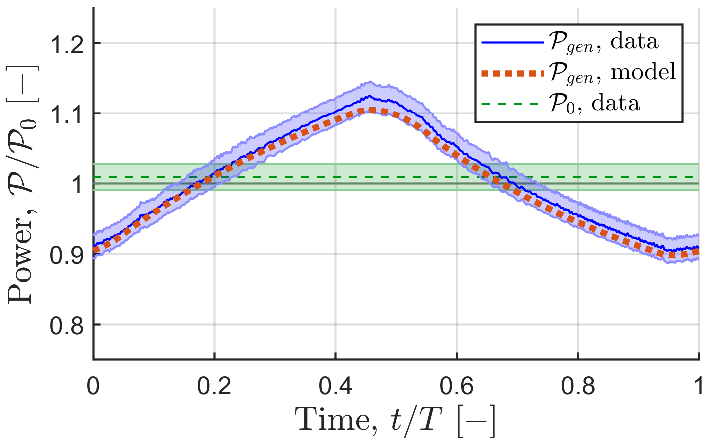}
		\caption{$A^* = 0.257$, $k = 0.455$, $u^* = 0.117$}
		\label{fig:trap25-1}
	\end{subfigure}
	\hfill
	\begin{subfigure}[t]{\columnwidth}
		\centering
		\includegraphics[width=\textwidth]{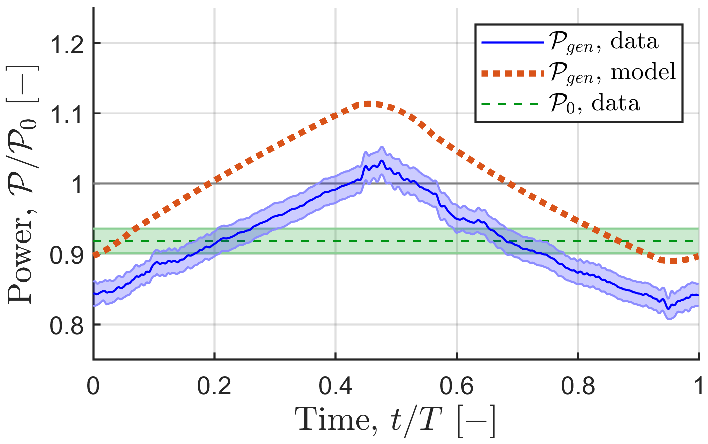}
		\caption{$A^* = 0.257$, $k = 0.911$, $u^* = 0.234$}
		\label{fig:trap25-2}
	\end{subfigure}
	\begin{subfigure}[t]{\columnwidth}
		\centering
		\includegraphics[width=\textwidth]{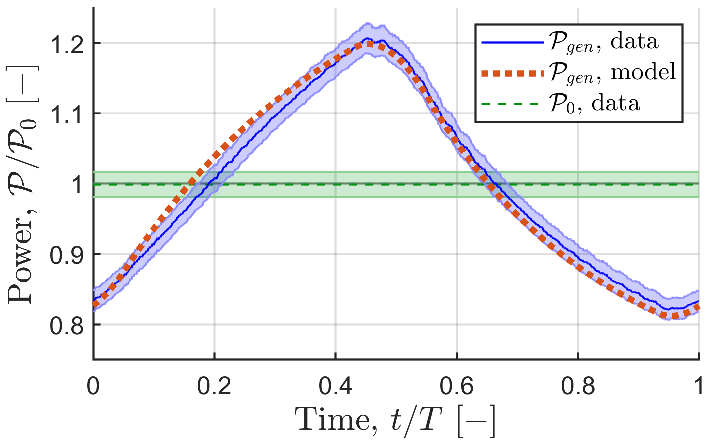}
		\caption{$A^* = 0.514$, $k = 0.304$, $u^* = 0.156$}
		\label{fig:trap25-3}
	\end{subfigure}
	\hfill
	\begin{subfigure}[t]{\columnwidth}
		\centering
		\includegraphics[width=\textwidth]{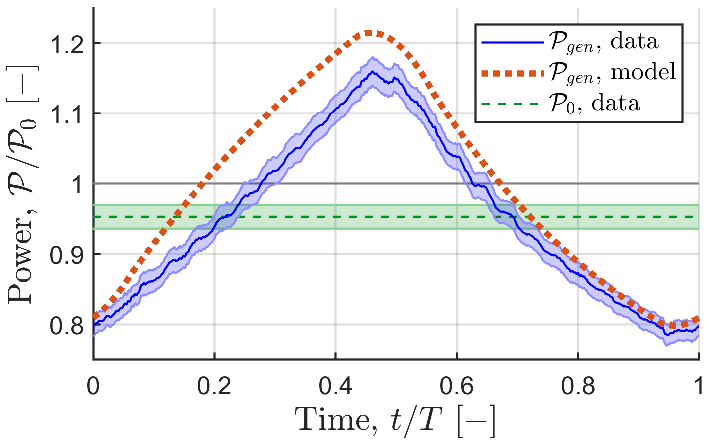}
		\caption{$A^* = 0.514$, $k = 0.455$, $u^* = 0.234$}
		\label{fig:trap25-4}
	\end{subfigure}
	\caption{Trapezoidal waveforms with $\xi=0.25$ and a load of 9.8 $\rm{\Omega}$ ($\lambda_0 = 6.11\pm0.25$), for four representative cases: (a) $A = 0.3$ m and $T = 2$ s, (b) $A = 0.3$ m and $T = 1$ s, (c) $A = 0.6$ m and $T = 3$ s, and (d) $A = 0.6$ m and $T = 2$ s.}
	\label{fig:trap25}
\end{figure*}

\begin{figure*}
	\begin{subfigure}[t]{\columnwidth}
		\centering
		\includegraphics[width=\textwidth]{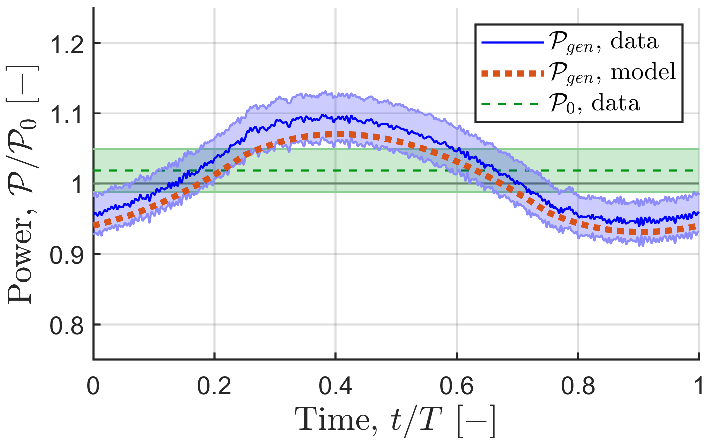}
		\caption{$A^* = 0.257$, $k = 0.455$, $u^* = 0.117$}
		\label{fig:trap100-1}
	\end{subfigure}
	\hfill
	\begin{subfigure}[t]{\columnwidth}
		\centering
		\includegraphics[width=\textwidth]{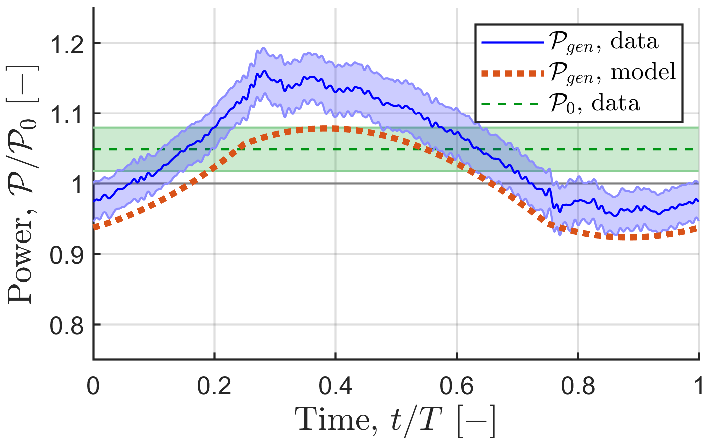}
		\caption{$A^* = 0.257$, $k = 0.911$, $u^* = 0.234$}
		\label{fig:trap100-2}
	\end{subfigure}
	\begin{subfigure}[t]{\columnwidth}
		\centering
		\includegraphics[width=\textwidth]{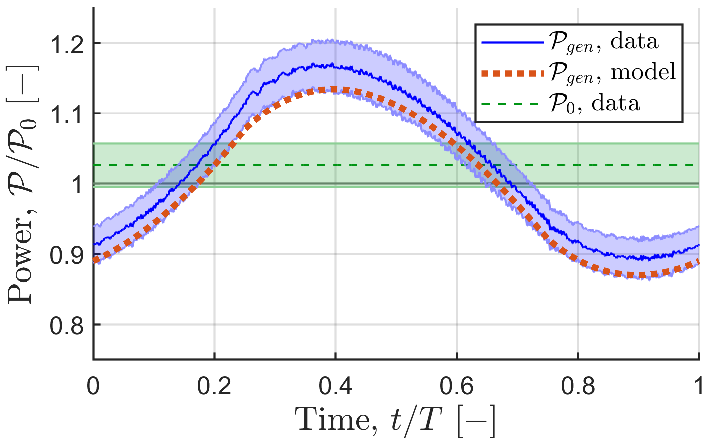}
		\caption{$A^* = 0.514$, $k = 0.304$, $u^* = 0.156$}
		\label{fig:trap100-3}
	\end{subfigure}
	\hfill
	\begin{subfigure}[t]{\columnwidth}
		\centering
		\includegraphics[width=\textwidth]{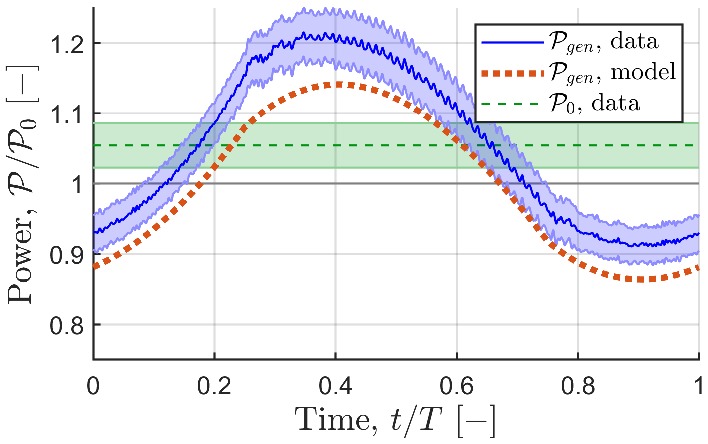}
		\caption{$A^* = 0.514$, $k = 0.455$, $u^* = 0.234$}
		\label{fig:trap100-4}
	\end{subfigure}
	\caption{Trapezoidal waveforms with $\xi=1$ and a load of 40 $\rm{\Omega}$ ($\lambda_0 = 8.77\pm0.35$), for four representative cases: (a) $A = 0.3$ m and $T = 2$ s, (b) $A = 0.3$ m and $T = 1$ s, (c) $A = 0.6$ m and $T = 3$ s, and (d) $A = 0.6$ m and $T = 2$ s.}
	\label{fig:trap100}
\end{figure*}

\begin{figure*}
	\begin{subfigure}[t]{\columnwidth}
		\centering
		\includegraphics[width=\textwidth]{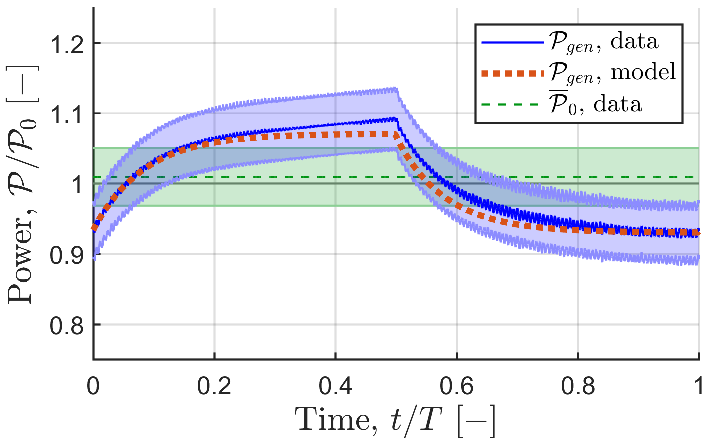}
		\caption{Load: 10 $\rm{\Omega}$ ($\lambda_0 = 6.27\pm0.26$)}
		\label{fig:trap01-1}
	\end{subfigure}
	\hfill
	\begin{subfigure}[t]{\columnwidth}
		\centering
		\includegraphics[width=\textwidth]{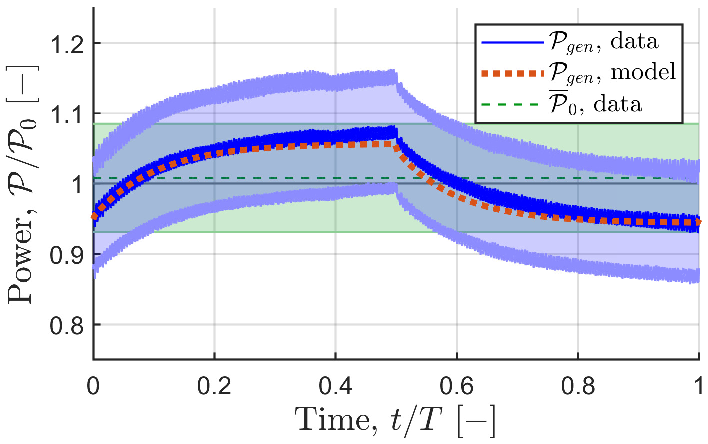}
		\caption{Load: 40 $\rm{\Omega}$ ($\lambda_0 = 8.77\pm0.35$)}
		\label{fig:trap01-2}
	\end{subfigure}
	\caption{Long-period trapezoidal waveforms with $\xi = 0.01$, $A^* = 0.257$, $k = 0.076$, and $u^* = 0.039$, for resistive loads of (a) 10 $\rm{\Omega}$ ($\lambda_0 = 6.27\pm0.26$) and (b) 40 $\rm{\Omega}$ ($\lambda_0 = 8.77\pm0.35$).}
	\label{fig:trap01}
\end{figure*}

A set of four cases, consisting of two nondimensional surge amplitudes ($A^* = 0.257$ and 0.514), three reduced frequencies ($k = 0.304$, 0.455, and 0.911), and three nondimensional surge velocities $u^* = 0.117$, 0.156, and 0.234) are shown for comparison across waveform shapes and loading conditions. This set of parameters is presented for various loading conditions and waveform types: sinusoidal waveforms with a load of 10 $\rm{\Omega}$ (Figure \ref{fig:sine10Ohm}), sinusoidal waveforms with a load of 40 $\rm{\Omega}$ (Figure \ref{fig:sine40Ohm}), trapezoidal waveforms with a load of 9.8 $\rm{\Omega}$ and $\xi=0.25$ (Figure \ref{fig:trap25}), and trapezoidal waveforms with a load of 40 $\rm{\Omega}$ and $\xi=1$ (Figure \ref{fig:trap100}). Additionally, to highlight the accuracy of the model in predicting time-resolved dynamics, two trapezoidal cases with $\xi=0.01$, $A^*=0.257$, $k=0.076$, and $u^*=0.039$ and loads of 10 and 40 $\rm{\Omega}$ are shown in Figure \ref{fig:trap01}. These cases are not hand-picked for agreement; they represent the general fidelity of the model with respect to the data in all of the cases presented in this study, with the exception of the cases involving the two lowest tip-speed ratios.

\bibliographystyle{apsrev4-1.bst}
\bibliography{SurgingTurbinePaper}

\begin{thebibliography}{31}%
\makeatletter
\providecommand \@ifxundefined [1]{%
 \@ifx{#1\undefined}
}%
\providecommand \@ifnum [1]{%
 \ifnum #1\expandafter \@firstoftwo
 \else \expandafter \@secondoftwo
 \fi
}%
\providecommand \@ifx [1]{%
 \ifx #1\expandafter \@firstoftwo
 \else \expandafter \@secondoftwo
 \fi
}%
\providecommand \natexlab [1]{#1}%
\providecommand \enquote  [1]{``#1''}%
\providecommand \bibnamefont  [1]{#1}%
\providecommand \bibfnamefont [1]{#1}%
\providecommand \citenamefont [1]{#1}%
\providecommand \href@noop [0]{\@secondoftwo}%
\providecommand \href [0]{\begingroup \@sanitize@url \@href}%
\providecommand \@href[1]{\@@startlink{#1}\@@href}%
\providecommand \@@href[1]{\endgroup#1\@@endlink}%
\providecommand \@sanitize@url [0]{\catcode `\\12\catcode `\$12\catcode
  `\&12\catcode `\#12\catcode `\^12\catcode `\_12\catcode `\%12\relax}%
\providecommand \@@startlink[1]{}%
\providecommand \@@endlink[0]{}%
\providecommand \url  [0]{\begingroup\@sanitize@url \@url }%
\providecommand \@url [1]{\endgroup\@href {#1}{\urlprefix }}%
\providecommand \urlprefix  [0]{URL }%
\providecommand \Eprint [0]{\href }%
\providecommand \doibase [0]{http://dx.doi.org/}%
\providecommand \selectlanguage [0]{\@gobble}%
\providecommand \bibinfo  [0]{\@secondoftwo}%
\providecommand \bibfield  [0]{\@secondoftwo}%
\providecommand \translation [1]{[#1]}%
\providecommand \BibitemOpen [0]{}%
\providecommand \bibitemStop [0]{}%
\providecommand \bibitemNoStop [0]{.\EOS\space}%
\providecommand \EOS [0]{\spacefactor3000\relax}%
\providecommand \BibitemShut  [1]{\csname bibitem#1\endcsname}%
\let\auto@bib@innerbib\@empty
\bibitem [{\citenamefont {Gueydon}\ \emph {et~al.}(2020)\citenamefont
  {Gueydon}, \citenamefont {Bayati},\ and\ \citenamefont
  {de~Ridder}}]{gueydon_discussion_2020}%
  \BibitemOpen
  \bibfield  {author} {\bibinfo {author} {\bibfnamefont {S.}~\bibnamefont
  {Gueydon}}, \bibinfo {author} {\bibfnamefont {I.}~\bibnamefont {Bayati}}, \
  and\ \bibinfo {author} {\bibfnamefont {E.}~\bibnamefont {de~Ridder}},\ }\href
  {\doibase 10.1016/j.oceaneng.2020.107288} {\bibfield  {journal} {\bibinfo
  {journal} {Ocean Engineering}\ }\textbf {\bibinfo {volume} {209}},\ \bibinfo
  {pages} {107288} (\bibinfo {year} {2020})}\BibitemShut {NoStop}%
\bibitem [{\citenamefont {Farrugia}\ \emph {et~al.}(2014)\citenamefont
  {Farrugia}, \citenamefont {Sant},\ and\ \citenamefont
  {Micallef}}]{farrugia_investigating_2014}%
  \BibitemOpen
  \bibfield  {author} {\bibinfo {author} {\bibfnamefont {R.}~\bibnamefont
  {Farrugia}}, \bibinfo {author} {\bibfnamefont {T.}~\bibnamefont {Sant}}, \
  and\ \bibinfo {author} {\bibfnamefont {D.}~\bibnamefont {Micallef}},\ }\href
  {\doibase 10.1016/j.renene.2013.12.043} {\bibfield  {journal} {\bibinfo
  {journal} {Renewable Energy}\ }\textbf {\bibinfo {volume} {70}},\ \bibinfo
  {pages} {24} (\bibinfo {year} {2014})}\BibitemShut {NoStop}%
\bibitem [{\citenamefont {Khosravi}\ \emph {et~al.}(2015)\citenamefont
  {Khosravi}, \citenamefont {Sarkar},\ and\ \citenamefont
  {Hu}}]{khosravi_experimental_2015}%
  \BibitemOpen
  \bibfield  {author} {\bibinfo {author} {\bibfnamefont {M.~M.}\ \bibnamefont
  {Khosravi}}, \bibinfo {author} {\bibfnamefont {P.}~\bibnamefont {Sarkar}}, \
  and\ \bibinfo {author} {\bibfnamefont {H.}~\bibnamefont {Hu}},\ }in\ \href
  {\doibase 10.2514/6.2015-1207} {\emph {\bibinfo {booktitle} {33rd {Wind}
  {Energy} {Symposium}}}}\ (\bibinfo  {publisher} {American Institute of
  Aeronautics and Astronautics},\ \bibinfo {year} {2015})\ \bibinfo {note}
  {\_eprint: https://arc.aiaa.org/doi/pdf/10.2514/6.2015-1207}\BibitemShut
  {NoStop}%
\bibitem [{\citenamefont {Shen}\ \emph {et~al.}(2018)\citenamefont {Shen},
  \citenamefont {Chen}, \citenamefont {Hu}, \citenamefont {Zhu},\ and\
  \citenamefont {Du}}]{shen_study_2018}%
  \BibitemOpen
  \bibfield  {author} {\bibinfo {author} {\bibfnamefont {X.}~\bibnamefont
  {Shen}}, \bibinfo {author} {\bibfnamefont {J.}~\bibnamefont {Chen}}, \bibinfo
  {author} {\bibfnamefont {P.}~\bibnamefont {Hu}}, \bibinfo {author}
  {\bibfnamefont {X.}~\bibnamefont {Zhu}}, \ and\ \bibinfo {author}
  {\bibfnamefont {Z.}~\bibnamefont {Du}},\ }\href {\doibase
  10.1016/j.energy.2017.12.100} {\bibfield  {journal} {\bibinfo  {journal}
  {Energy}\ }\textbf {\bibinfo {volume} {145}},\ \bibinfo {pages} {793}
  (\bibinfo {year} {2018})}\BibitemShut {NoStop}%
\bibitem [{\citenamefont {Farrugia}\ \emph {et~al.}(2016)\citenamefont
  {Farrugia}, \citenamefont {Sant},\ and\ \citenamefont
  {Micallef}}]{farrugia_study_2016}%
  \BibitemOpen
  \bibfield  {author} {\bibinfo {author} {\bibfnamefont {R.}~\bibnamefont
  {Farrugia}}, \bibinfo {author} {\bibfnamefont {T.}~\bibnamefont {Sant}}, \
  and\ \bibinfo {author} {\bibfnamefont {D.}~\bibnamefont {Micallef}},\ }\href
  {\doibase 10.1016/j.renene.2015.08.063} {\bibfield  {journal} {\bibinfo
  {journal} {Renewable Energy}\ }\textbf {\bibinfo {volume} {86}},\ \bibinfo
  {pages} {770} (\bibinfo {year} {2016})}\BibitemShut {NoStop}%
\bibitem [{\citenamefont {Wen}\ \emph {et~al.}(2017)\citenamefont {Wen},
  \citenamefont {Tian}, \citenamefont {Dong}, \citenamefont {Peng},\ and\
  \citenamefont {Zhang}}]{wen_influences_2017}%
  \BibitemOpen
  \bibfield  {author} {\bibinfo {author} {\bibfnamefont {B.}~\bibnamefont
  {Wen}}, \bibinfo {author} {\bibfnamefont {X.}~\bibnamefont {Tian}}, \bibinfo
  {author} {\bibfnamefont {X.}~\bibnamefont {Dong}}, \bibinfo {author}
  {\bibfnamefont {Z.}~\bibnamefont {Peng}}, \ and\ \bibinfo {author}
  {\bibfnamefont {W.}~\bibnamefont {Zhang}},\ }\href {\doibase
  10.1016/j.energy.2017.11.090} {\bibfield  {journal} {\bibinfo  {journal}
  {Energy}\ }\textbf {\bibinfo {volume} {141}},\ \bibinfo {pages} {2054}
  (\bibinfo {year} {2017})}\BibitemShut {NoStop}%
\bibitem [{\citenamefont {Johlas}\ \emph {et~al.}(2021)\citenamefont {Johlas},
  \citenamefont {Mart{\'i}nez-Tossas}, \citenamefont {Churchfield},
  \citenamefont {Lackner},\ and\ \citenamefont
  {Schmidt}}]{johlas_floating_2021}%
  \BibitemOpen
  \bibfield  {author} {\bibinfo {author} {\bibfnamefont {H.~M.}\ \bibnamefont
  {Johlas}}, \bibinfo {author} {\bibfnamefont {L.~A.}\ \bibnamefont
  {Mart{\'i}nez-Tossas}}, \bibinfo {author} {\bibfnamefont {M.~J.}\
  \bibnamefont {Churchfield}}, \bibinfo {author} {\bibfnamefont {M.~A.}\
  \bibnamefont {Lackner}}, \ and\ \bibinfo {author} {\bibfnamefont {D.~P.}\
  \bibnamefont {Schmidt}},\ }\href {\doibase 10.1002/we.2608} {\bibfield
  {journal} {\bibinfo  {journal} {Wind Energy}\ }\textbf {\bibinfo {volume}
  {24}},\ \bibinfo {pages} {901} (\bibinfo {year} {2021})},\ \bibinfo {note}
  {\_eprint:
  https://onlinelibrary.wiley.com/doi/pdf/10.1002/we.2608}\BibitemShut
  {NoStop}%
\bibitem [{\citenamefont {Tran}\ and\ \citenamefont
  {Kim}(2016)}]{tran_cfd_2016}%
  \BibitemOpen
  \bibfield  {author} {\bibinfo {author} {\bibfnamefont {T.~T.}\ \bibnamefont
  {Tran}}\ and\ \bibinfo {author} {\bibfnamefont {D.-H.}\ \bibnamefont {Kim}},\
  }\href {\doibase 10.1016/j.renene.2015.12.013} {\bibfield  {journal}
  {\bibinfo  {journal} {Renewable Energy}\ }\textbf {\bibinfo {volume} {90}},\
  \bibinfo {pages} {204} (\bibinfo {year} {2016})}\BibitemShut {NoStop}%
\bibitem [{\citenamefont {Mancini}\ \emph {et~al.}(2020)\citenamefont
  {Mancini}, \citenamefont {Boorsma}, \citenamefont {Caboni}, \citenamefont
  {Cormier}, \citenamefont {Lutz}, \citenamefont {Schito},\ and\ \citenamefont
  {Zasso}}]{mancini_characterization_2020}%
  \BibitemOpen
  \bibfield  {author} {\bibinfo {author} {\bibfnamefont {S.}~\bibnamefont
  {Mancini}}, \bibinfo {author} {\bibfnamefont {K.}~\bibnamefont {Boorsma}},
  \bibinfo {author} {\bibfnamefont {M.}~\bibnamefont {Caboni}}, \bibinfo
  {author} {\bibfnamefont {M.}~\bibnamefont {Cormier}}, \bibinfo {author}
  {\bibfnamefont {T.}~\bibnamefont {Lutz}}, \bibinfo {author} {\bibfnamefont
  {P.}~\bibnamefont {Schito}}, \ and\ \bibinfo {author} {\bibfnamefont
  {A.}~\bibnamefont {Zasso}},\ }\href {\doibase 10.5194/wes-5-1713-2020}
  {\bibfield  {journal} {\bibinfo  {journal} {Wind Energy Science}\ }\textbf
  {\bibinfo {volume} {5}},\ \bibinfo {pages} {1713} (\bibinfo {year} {2020})},\
  \bibinfo {note} {publisher: Copernicus GmbH}\BibitemShut {NoStop}%
\bibitem [{\citenamefont {Sant}\ \emph {et~al.}(2015)\citenamefont {Sant},
  \citenamefont {Bonnici}, \citenamefont {Farrugia},\ and\ \citenamefont
  {Micallef}}]{sant_measurements_2015}%
  \BibitemOpen
  \bibfield  {author} {\bibinfo {author} {\bibfnamefont {T.}~\bibnamefont
  {Sant}}, \bibinfo {author} {\bibfnamefont {D.}~\bibnamefont {Bonnici}},
  \bibinfo {author} {\bibfnamefont {R.}~\bibnamefont {Farrugia}}, \ and\
  \bibinfo {author} {\bibfnamefont {D.}~\bibnamefont {Micallef}},\ }\href
  {\doibase 10.1002/we.1730} {\bibfield  {journal} {\bibinfo  {journal} {Wind
  Energy}\ }\textbf {\bibinfo {volume} {18}},\ \bibinfo {pages} {811} (\bibinfo
  {year} {2015})},\ \bibinfo {note} {\_eprint:
  https://onlinelibrary.wiley.com/doi/pdf/10.1002/we.1730}\BibitemShut
  {NoStop}%
\bibitem [{\citenamefont {Wen}\ \emph {et~al.}(2018{\natexlab{a}})\citenamefont
  {Wen}, \citenamefont {Tian}, \citenamefont {Dong}, \citenamefont {Peng},\
  and\ \citenamefont {Zhang}}]{wen_power_2018}%
  \BibitemOpen
  \bibfield  {author} {\bibinfo {author} {\bibfnamefont {B.}~\bibnamefont
  {Wen}}, \bibinfo {author} {\bibfnamefont {X.}~\bibnamefont {Tian}}, \bibinfo
  {author} {\bibfnamefont {X.}~\bibnamefont {Dong}}, \bibinfo {author}
  {\bibfnamefont {Z.}~\bibnamefont {Peng}}, \ and\ \bibinfo {author}
  {\bibfnamefont {W.}~\bibnamefont {Zhang}},\ }\href {\doibase 10.1002/we.2215}
  {\bibfield  {journal} {\bibinfo  {journal} {Wind Energy}\ }\textbf {\bibinfo
  {volume} {21}},\ \bibinfo {pages} {1076} (\bibinfo {year}
  {2018}{\natexlab{a}})},\ \bibinfo {note} {\_eprint:
  https://onlinelibrary.wiley.com/doi/pdf/10.1002/we.2215}\BibitemShut
  {NoStop}%
\bibitem [{\citenamefont {Micallef}\ and\ \citenamefont
  {Sant}(2015)}]{micallef_loading_2015}%
  \BibitemOpen
  \bibfield  {author} {\bibinfo {author} {\bibfnamefont {D.}~\bibnamefont
  {Micallef}}\ and\ \bibinfo {author} {\bibfnamefont {T.}~\bibnamefont
  {Sant}},\ }\href {\doibase 10.1016/j.renene.2015.05.016} {\bibfield
  {journal} {\bibinfo  {journal} {Renewable Energy}\ }\textbf {\bibinfo
  {volume} {83}},\ \bibinfo {pages} {737} (\bibinfo {year} {2015})}\BibitemShut
  {NoStop}%
\bibitem [{\citenamefont {Wen}\ \emph {et~al.}(2018{\natexlab{b}})\citenamefont
  {Wen}, \citenamefont {Dong}, \citenamefont {Tian}, \citenamefont {Peng},
  \citenamefont {Zhang},\ and\ \citenamefont {Wei}}]{wen_power_2018-1}%
  \BibitemOpen
  \bibfield  {author} {\bibinfo {author} {\bibfnamefont {B.}~\bibnamefont
  {Wen}}, \bibinfo {author} {\bibfnamefont {X.}~\bibnamefont {Dong}}, \bibinfo
  {author} {\bibfnamefont {X.}~\bibnamefont {Tian}}, \bibinfo {author}
  {\bibfnamefont {Z.}~\bibnamefont {Peng}}, \bibinfo {author} {\bibfnamefont
  {W.}~\bibnamefont {Zhang}}, \ and\ \bibinfo {author} {\bibfnamefont
  {K.}~\bibnamefont {Wei}},\ }\href {\doibase 10.1016/j.energy.2018.04.140}
  {\bibfield  {journal} {\bibinfo  {journal} {Energy}\ }\textbf {\bibinfo
  {volume} {154}},\ \bibinfo {pages} {508} (\bibinfo {year}
  {2018}{\natexlab{b}})}\BibitemShut {NoStop}%
\bibitem [{\citenamefont {de~Vaal}\ \emph {et~al.}(2014)\citenamefont
  {de~Vaal}, \citenamefont {Hansen},\ and\ \citenamefont
  {Moan}}]{de_vaal_effect_2014}%
  \BibitemOpen
  \bibfield  {author} {\bibinfo {author} {\bibfnamefont {J.~B.}\ \bibnamefont
  {de~Vaal}}, \bibinfo {author} {\bibfnamefont {M.~O.~L.}\ \bibnamefont
  {Hansen}}, \ and\ \bibinfo {author} {\bibfnamefont {T.}~\bibnamefont
  {Moan}},\ }\href {\doibase 10.1002/we.1562} {\bibfield  {journal} {\bibinfo
  {journal} {Wind Energy}\ }\textbf {\bibinfo {volume} {17}},\ \bibinfo {pages}
  {105} (\bibinfo {year} {2014})},\ \bibinfo {note} {\_eprint:
  https://onlinelibrary.wiley.com/doi/pdf/10.1002/we.1562}\BibitemShut
  {NoStop}%
\bibitem [{\citenamefont {Fontanella}\ \emph {et~al.}(2020)\citenamefont
  {Fontanella}, \citenamefont {Al}, \citenamefont {Hoek}, \citenamefont {Liu},
  \citenamefont {Wingerden},\ and\ \citenamefont
  {Belloli}}]{fontanella_control-oriented_2020}%
  \BibitemOpen
  \bibfield  {author} {\bibinfo {author} {\bibfnamefont {A.}~\bibnamefont
  {Fontanella}}, \bibinfo {author} {\bibfnamefont {M.}~\bibnamefont {Al}},
  \bibinfo {author} {\bibfnamefont {D.~v.~d.}\ \bibnamefont {Hoek}}, \bibinfo
  {author} {\bibfnamefont {Y.}~\bibnamefont {Liu}}, \bibinfo {author}
  {\bibfnamefont {J.~W.~v.}\ \bibnamefont {Wingerden}}, \ and\ \bibinfo
  {author} {\bibfnamefont {M.}~\bibnamefont {Belloli}},\ }\href {\doibase
  10.1088/1742-6596/1618/2/022038} {\bibfield  {journal} {\bibinfo  {journal}
  {Journal of Physics: Conference Series}\ }\textbf {\bibinfo {volume}
  {1618}},\ \bibinfo {pages} {022038} (\bibinfo {year} {2020})},\ \bibinfo
  {note} {publisher: IOP Publishing}\BibitemShut {NoStop}%
\bibitem [{\citenamefont {Fontanella}\ \emph {et~al.}(2021)\citenamefont
  {Fontanella}, \citenamefont {Al}, \citenamefont {van Wingerden},\ and\
  \citenamefont {Belloli}}]{fontanella_model-based_2021}%
  \BibitemOpen
  \bibfield  {author} {\bibinfo {author} {\bibfnamefont {A.}~\bibnamefont
  {Fontanella}}, \bibinfo {author} {\bibfnamefont {M.}~\bibnamefont {Al}},
  \bibinfo {author} {\bibfnamefont {J.-W.}\ \bibnamefont {van Wingerden}}, \
  and\ \bibinfo {author} {\bibfnamefont {M.}~\bibnamefont {Belloli}},\ }\href
  {\doibase 10.5194/wes-6-885-2021} {\bibfield  {journal} {\bibinfo  {journal}
  {Wind Energy Science}\ }\textbf {\bibinfo {volume} {6}},\ \bibinfo {pages}
  {885} (\bibinfo {year} {2021})},\ \bibinfo {note} {publisher: Copernicus
  GmbH}\BibitemShut {NoStop}%
\bibitem [{\citenamefont {Sebastian}\ and\ \citenamefont
  {Lackner}(2013)}]{sebastian_characterization_2013}%
  \BibitemOpen
  \bibfield  {author} {\bibinfo {author} {\bibfnamefont {T.}~\bibnamefont
  {Sebastian}}\ and\ \bibinfo {author} {\bibfnamefont {M.~A.}\ \bibnamefont
  {Lackner}},\ }\href {\doibase 10.1002/we.545} {\bibfield  {journal} {\bibinfo
   {journal} {Wind Energy}\ }\textbf {\bibinfo {volume} {16}},\ \bibinfo
  {pages} {339} (\bibinfo {year} {2013})}\BibitemShut {NoStop}%
\bibitem [{\citenamefont {Rockel}\ \emph {et~al.}(2016)\citenamefont {Rockel},
  \citenamefont {Peinke}, \citenamefont {H{\"o}lling},\ and\ \citenamefont
  {Cal}}]{rockel_wake_2016}%
  \BibitemOpen
  \bibfield  {author} {\bibinfo {author} {\bibfnamefont {S.}~\bibnamefont
  {Rockel}}, \bibinfo {author} {\bibfnamefont {J.}~\bibnamefont {Peinke}},
  \bibinfo {author} {\bibfnamefont {M.}~\bibnamefont {H{\"o}lling}}, \ and\
  \bibinfo {author} {\bibfnamefont {R.~B.}\ \bibnamefont {Cal}},\ }\href
  {\doibase 10.1016/j.renene.2015.07.012} {\bibfield  {journal} {\bibinfo
  {journal} {Renewable Energy}\ }\textbf {\bibinfo {volume} {85}},\ \bibinfo
  {pages} {666} (\bibinfo {year} {2016})}\BibitemShut {NoStop}%
\bibitem [{\citenamefont {Bayati}\ \emph {et~al.}(2017)\citenamefont {Bayati},
  \citenamefont {Belloli}, \citenamefont {Bernini},\ and\ \citenamefont
  {Zasso}}]{bayati_wind_2017}%
  \BibitemOpen
  \bibfield  {author} {\bibinfo {author} {\bibfnamefont {I.}~\bibnamefont
  {Bayati}}, \bibinfo {author} {\bibfnamefont {M.}~\bibnamefont {Belloli}},
  \bibinfo {author} {\bibfnamefont {L.}~\bibnamefont {Bernini}}, \ and\
  \bibinfo {author} {\bibfnamefont {A.}~\bibnamefont {Zasso}},\ }\href
  {\doibase 10.1016/j.egypro.2017.10.375} {\bibfield  {journal} {\bibinfo
  {journal} {Energy Procedia}\ }\bibinfo {series} {14th {Deep} {Sea} {Offshore}
  {Wind} {R}\&{D} {Conference}, {EERA} {DeepWind}'2017},\ \textbf {\bibinfo
  {volume} {137}},\ \bibinfo {pages} {214} (\bibinfo {year}
  {2017})}\BibitemShut {NoStop}%
\bibitem [{\citenamefont {Lee}\ and\ \citenamefont
  {Lee}(2019)}]{lee_effects_2019}%
  \BibitemOpen
  \bibfield  {author} {\bibinfo {author} {\bibfnamefont {H.}~\bibnamefont
  {Lee}}\ and\ \bibinfo {author} {\bibfnamefont {D.-J.}\ \bibnamefont {Lee}},\
  }\href {\doibase 10.1016/j.renene.2019.04.134} {\bibfield  {journal}
  {\bibinfo  {journal} {Renewable Energy}\ }\textbf {\bibinfo {volume} {143}},\
  \bibinfo {pages} {9} (\bibinfo {year} {2019})}\BibitemShut {NoStop}%
\bibitem [{\citenamefont {Kopperstad}\ \emph {et~al.}(2020)\citenamefont
  {Kopperstad}, \citenamefont {Kumar},\ and\ \citenamefont
  {Shoele}}]{kopperstad_aerodynamic_2020}%
  \BibitemOpen
  \bibfield  {author} {\bibinfo {author} {\bibfnamefont {K.~M.}\ \bibnamefont
  {Kopperstad}}, \bibinfo {author} {\bibfnamefont {R.}~\bibnamefont {Kumar}}, \
  and\ \bibinfo {author} {\bibfnamefont {K.}~\bibnamefont {Shoele}},\ }\href
  {\doibase 10.1002/we.2547} {\bibfield  {journal} {\bibinfo  {journal} {Wind
  Energy}\ }\textbf {\bibinfo {volume} {23}},\ \bibinfo {pages} {2087}
  (\bibinfo {year} {2020})},\ \bibinfo {note} {\_eprint:
  https://onlinelibrary.wiley.com/doi/pdf/10.1002/we.2547}\BibitemShut
  {NoStop}%
\bibitem [{\citenamefont {Rezaeiha}\ and\ \citenamefont
  {Micallef}(2021)}]{rezaeiha_wake_2021}%
  \BibitemOpen
  \bibfield  {author} {\bibinfo {author} {\bibfnamefont {A.}~\bibnamefont
  {Rezaeiha}}\ and\ \bibinfo {author} {\bibfnamefont {D.}~\bibnamefont
  {Micallef}},\ }\href {\doibase 10.1016/j.renene.2021.07.087} {\bibfield
  {journal} {\bibinfo  {journal} {Renewable Energy}\ }\textbf {\bibinfo
  {volume} {179}},\ \bibinfo {pages} {859} (\bibinfo {year}
  {2021})}\BibitemShut {NoStop}%
\bibitem [{\citenamefont {Cherubini}\ \emph {et~al.}(2015)\citenamefont
  {Cherubini}, \citenamefont {Papini}, \citenamefont {Vertechy},\ and\
  \citenamefont {Fontana}}]{cherubini_airborne_2015}%
  \BibitemOpen
  \bibfield  {author} {\bibinfo {author} {\bibfnamefont {A.}~\bibnamefont
  {Cherubini}}, \bibinfo {author} {\bibfnamefont {A.}~\bibnamefont {Papini}},
  \bibinfo {author} {\bibfnamefont {R.}~\bibnamefont {Vertechy}}, \ and\
  \bibinfo {author} {\bibfnamefont {M.}~\bibnamefont {Fontana}},\ }\href
  {\doibase 10.1016/j.rser.2015.07.053} {\bibfield  {journal} {\bibinfo
  {journal} {Renewable and Sustainable Energy Reviews}\ }\textbf {\bibinfo
  {volume} {51}},\ \bibinfo {pages} {1461} (\bibinfo {year}
  {2015})}\BibitemShut {NoStop}%
\bibitem [{\citenamefont {Jonkman}(2021)}]{jonkman_google_2021}%
  \BibitemOpen
  \bibfield  {author} {\bibinfo {author} {\bibfnamefont {J.}~\bibnamefont
  {Jonkman}},\ }\href {\doibase 10.2172/1813012} {\emph {\bibinfo {title}
  {Google / {Makani} {Energy} {Kite} {Modeling}: {Cooperative} {Research} and
  {Development} {Final} {Report}, {CRADA} {Number} {CRD}-17-00569}}},\ \bibinfo
  {type} {Tech. Rep.}\ \bibinfo {number} {NREL/TP-5000-80635}\ (\bibinfo
  {institution} {National Renewable Energy Lab. (NREL), Golden, CO (United
  States)},\ \bibinfo {year} {2021})\BibitemShut {NoStop}%
\bibitem [{\citenamefont {Dabiri}(2020)}]{dabiri_theoretical_2020}%
  \BibitemOpen
  \bibfield  {author} {\bibinfo {author} {\bibfnamefont {J.~O.}\ \bibnamefont
  {Dabiri}},\ }\href {\doibase 10.1103/PhysRevFluids.5.022501} {\bibfield
  {journal} {\bibinfo  {journal} {Physical Review Fluids}\ }\textbf {\bibinfo
  {volume} {5}},\ \bibinfo {pages} {022501} (\bibinfo {year} {2020})},\
  \bibinfo {note} {publisher: American Physical Society}\BibitemShut {NoStop}%
\bibitem [{\citenamefont {Concordia}(1951)}]{concordia_synchronous_1951}%
  \BibitemOpen
  \bibfield  {author} {\bibinfo {author} {\bibfnamefont {C.}~\bibnamefont
  {Concordia}},\ }\href {https://books.google.com/books?id=5wgjAAAAMAAJ} {\emph
  {\bibinfo {title} {Synchronous {Machines}: {Theory} and {Performance}}}},\
  General {Electric} series\ (\bibinfo  {publisher} {Wiley},\ \bibinfo {year}
  {1951})\BibitemShut {NoStop}%
\bibitem [{\citenamefont {Stevenson}\ and\ \citenamefont
  {Grainger}(1994)}]{stevenson_power_1994}%
  \BibitemOpen
  \bibfield  {author} {\bibinfo {author} {\bibfnamefont {W.}~\bibnamefont
  {Stevenson}}\ and\ \bibinfo {author} {\bibfnamefont {J.}~\bibnamefont
  {Grainger}},\ }\href {https://books.google.com/books?id=NBIoAQAAMAAJ} {\emph
  {\bibinfo {title} {Power {System} {Analysis}}}}\ (\bibinfo  {publisher}
  {McGraw-Hill Education},\ \bibinfo {year} {1994})\BibitemShut {NoStop}%
\bibitem [{\citenamefont {Sears}(1941)}]{sears_aspects_1941}%
  \BibitemOpen
  \bibfield  {author} {\bibinfo {author} {\bibfnamefont {W.~R.}\ \bibnamefont
  {Sears}},\ }\href {\doibase 10.2514/8.10655} {\bibfield  {journal} {\bibinfo
  {journal} {Journal of the Aeronautical Sciences}\ }\textbf {\bibinfo {volume}
  {8}},\ \bibinfo {pages} {104} (\bibinfo {year} {1941})}\BibitemShut {NoStop}%
\bibitem [{\citenamefont {El~Makdah}\ \emph {et~al.}(2019)\citenamefont
  {El~Makdah}, \citenamefont {Ruzzante}, \citenamefont {Zhang},\ and\
  \citenamefont {Rival}}]{el_makdah_influence_2019}%
  \BibitemOpen
  \bibfield  {author} {\bibinfo {author} {\bibfnamefont {A.~M.}\ \bibnamefont
  {El~Makdah}}, \bibinfo {author} {\bibfnamefont {S.}~\bibnamefont {Ruzzante}},
  \bibinfo {author} {\bibfnamefont {K.}~\bibnamefont {Zhang}}, \ and\ \bibinfo
  {author} {\bibfnamefont {D.~E.}\ \bibnamefont {Rival}},\ }\href {\doibase
  10.1016/j.jfluidstructs.2019.04.009} {\bibfield  {journal} {\bibinfo
  {journal} {Journal of Fluids and Structures}\ }\textbf {\bibinfo {volume}
  {88}},\ \bibinfo {pages} {71} (\bibinfo {year} {2019})}\BibitemShut {NoStop}%
\bibitem [{\citenamefont {Miller}\ \emph {et~al.}(2019)\citenamefont {Miller},
  \citenamefont {Kiefer}, \citenamefont {Westergaard}, \citenamefont {Hansen},\
  and\ \citenamefont {Hultmark}}]{miller_horizontal_2019}%
  \BibitemOpen
  \bibfield  {author} {\bibinfo {author} {\bibfnamefont {M.~A.}\ \bibnamefont
  {Miller}}, \bibinfo {author} {\bibfnamefont {J.}~\bibnamefont {Kiefer}},
  \bibinfo {author} {\bibfnamefont {C.}~\bibnamefont {Westergaard}}, \bibinfo
  {author} {\bibfnamefont {M.~O.~L.}\ \bibnamefont {Hansen}}, \ and\ \bibinfo
  {author} {\bibfnamefont {M.}~\bibnamefont {Hultmark}},\ }\href {\doibase
  10.1103/PhysRevFluids.4.110504} {\bibfield  {journal} {\bibinfo  {journal}
  {Physical Review Fluids}\ }\textbf {\bibinfo {volume} {4}},\ \bibinfo {pages}
  {110504} (\bibinfo {year} {2019})},\ \bibinfo {note} {publisher: American
  Physical Society}\BibitemShut {NoStop}%
\bibitem [{\citenamefont {Betz}(1920)}]{betz_maximum_1920}%
  \BibitemOpen
  \bibfield  {author} {\bibinfo {author} {\bibfnamefont {A.}~\bibnamefont
  {Betz}},\ }\href@noop {} {\bibfield  {journal} {\bibinfo  {journal}
  {Zeitschrift f{\"u}r das gesamte Turbinenwesen}\ }\textbf {\bibinfo {volume}
  {26}},\ \bibinfo {pages} {307} (\bibinfo {year} {1920})}\BibitemShut
  {NoStop}%
\end{thebibliography}%

\end{document}